\documentclass{aa}
\usepackage[usenames,dvipsnames]{xcolor}
\usepackage{natbib}
\usepackage{amssymb}
\usepackage{amsmath}
\usepackage{graphicx}
\usepackage{hyperref}
\usepackage[T1]{fontenc}
\definecolor{revcol}{RGB}{216, 18, 125}
\def\revised#1{#1}
\def\revisedagain#1{#1}

\newcommand{\removed}[1]{}

\definecolor{intrevcolor}{rgb}{0.0, 0.0, 1}
\definecolor{newstuffcolor}{rgb}{0.0, 0.6, 0}
\begin{document}

\title{Midplane temperature and outer edge of the protoplanetary disk around HD 163296}
\titlerunning{Midplane temperature and outer edge of a protoplanetary disk}
\authorrunning{Dullemond et al.}

\author{C.P.~Dullemond$^{1}$, A.~Isella$^{2}$, S.M.~Andrews$^{3}$, I.~Skobleva$^{1}$, N.~Dzyurkevich$^{1}$}
\institute{
  (1) Zentrum f\"ur Astronomie, Heidelberg University, Albert Ueberle Str.~2, 69120 Heidelberg, Germany\\
  (2) Department of Physics and Astronomy, Rice University 6100 Main Street, MS-108, Houston, TX 77005, USA\\
  (3) Harvard-Smithsonian Center for Astrophysics, 60 Garden Street, Cambridge, MA 02138, USA\\
} \date{\today}

\abstract{ 
  Knowledge of the midplane temperature of protoplanetary disks is one of the
  key ingredients in theories of dust growth and planet formation. However,
  direct measurement of this quantity is complicated, and often depends on the
  fitting of complex models to the data. In this paper we demonstrate a method
  to directly measure the midplane gas temperature from an optically thick
  molecular line, if the disk is moderately inclined. The only model assumption
  that enters is that the line is very optically thick, also in the midplane
  region where we wish to measure the temperature. Freeze-out of the molecule
  onto dust grains could thwart this. However, in regions that are expected to
  be warm enough to avoid freeze-out, this method should work. We apply the
  method to the CO 2-1 line channel maps of the disk around HD 163296. We find
  that the midplane temperature between 100 and 400 au drops only mildly from 25
  K down to 18 K. \revised{While we see no direct evidence of the midplane being
    optically thin due to strong CO depletion by freeze-out, we cannot rule it out
    either. The fact that the inferred temperatures are close to the expected
    CO freeze-out temperature could be an indication of this.}
  Incidently, for the disk around HD 163296 we also find dynamic evidence for a rather abrupt
  outer edge of the disk, suggestive of outside-in photoevaporation or
  truncation by an unseen companion.
}

\maketitle


\section{Introduction}
One of the prime goals of observations of protoplanetary disks is to improve our
understanding of the environment in which planets form. Measurements of the
density and temperature distribution in these disks are of fundamental
importance. Spatially resolved observations of rotational lines allow the
measurement of the gas temperature, but there are caveats. For instance: while
it is relatively straightforward to measure the gas temperature in the disk's
surface layers, it is much harder to do that in the disk's cold midplane region.

Protoplanetary disks are known to have a vertical temperature gradient: The
surface layers are warmer than the midplane regions
\citep[e.g.][]{1998ApJ...500..411D}. The temperature in these warm surface
layers can be probed with optically thick molecular lines such as the $J=$ 2-1
CO line, which is readily accessible with ALMA. As long as the level populations
of the molecule are in local thermodynamic equilibrium (LTE), and the CO line is
optically thick, the brightness temperature of the CO emission at the local line
center is a direct measure of the gas temperature in that layer
\citep{2018ApJ...853..113W}. This is, in fact, an application of the
Eddington-Barbier rule for LTE radiative transfer: The observed intensity
$I_\nu$ is roughly equal to the Planck function $B_\nu(T)$ at the depth in the
photosphere where $\tau_\nu\simeq 1$. For highly optically thick lines, the
$\tau_\nu\simeq 1$ surface of a protoplanetary disk at the local line center is
located high above the midplane. The inferred gas temperature is thus that of
the warm surface layers of the disk, not that of the cold midplane.

However, if one is interested in the conditions under which dust growth and
planet formation take place, the relevant temperature is that of the much
cooler midplane. The high optical depth of a CO line typically shields the
midplane regions from view. To probe these regions one could resort to optically
thin molecular lines. However, for optically thin lines Eddington-Barbier's rule
cannot be applied, and the inference of the gas temperature from such optically
thin lines becomes much more model-dependent and insecure.

The problem of high optical depth of a line that shields the midplane can, however,
be overcome for non-face-on disks, as was shown by \citet{2013ApJ...774...16R},
and as we shall further demonstrate in this paper.

The trick is to make use of the morphology of the line emission channel maps of
a keplerianly rotating disk, when seen at a moderate inclination. Such maps show
one-sided teacup-handle shaped emission at the location where the projected
Kepler velocity equals the velocity-offset of the channel. As shown by radiative
transfer models \citep[e.g.][]{2007ApJ...669.1262P}, the vertical temperature
structure tends to split this ``teacup handle'' into two: one from the upper
warm surface layer, and one from the lower warm surface layer. The cold midplane
does not emit much. This is in part due to the possibility that the CO gas is
frozen out onto dust grains in these regions. But even if no freeze-out occurs,
the lower temperature of the gas in the midplane regions simply causes the gas
to emit less radiation. The split ``teacup handle'' thus gets the shape of the
double wings of a dragonfly, albeit one-sided.

One of the protoplanetary disks for which this ``dragonfly wings'' geometry of CO
rotational line channel maps is most clearly seen is HD 163296
\citep{2013ApJ...774...16R, 2018ApJ...860L..13P, 2018ApJ...869L..49I}. The CO
2-1 maps from the DSHARP campaign \citep{2018ApJ...869L..49I} have sufficient
signal-to-noise that even the weak thermal emission from between the two
dragonfly wings is measurable. This is the key to the method described in this
paper.

In this paper we will demonstrate that if the observations show non-zero
brightness between the upper and lower ``wings'' of the dragonfly, this can be
directly translated, using the Eddington-Barbier rule, into the midplane gas
temperature. This measurement is direct: it does not require radiative transfer
modeling to be interpreted. We will apply this to the CO channel maps of the
source HD 163296.

\section{Testing the method using radiative transfer}
\label{sec-radtrans-model}

\subsection{Model setup}
To demonstrate the principle, we set up a radiative transfer model with the
{\small\tt RADMC-3D}
code\footnote{\url{http://www.ita.uni-heidelberg.de/~dullemond/software/radmc-3d/}}
\citep{2012ascl.soft02015D}
aimed to qualitatively reproduce the CO $J=$ 2-1 channel maps of HD 163296 from
the DSHARP campaign \citep{2018ApJ...869L..49I}. We do not aim to make an exact
replica of the disk. The main aim is to demonstrate the radiative transfer
effects on the channel maps, so as to learn how to ``read'' them.

The parameters of the star are taken from \citet{2018ApJ...869L..41A}:
$M_{*}=2.04\,M_{\odot}$, $L_{*}=17\,L_{\odot}$, $T_{*}=9332\,\mathrm{K}$, and
the distance is $101\,\mathrm{pc}$. The disk has an inclination with the line of
sight of $i=46.7^{\circ{}}$, and a position angle of $133.3^{\circ{}}$. We
assume the disk is axially symmetric. We take the disk gas surface density to be
\begin{equation}\label{eq-surf-dens-model}
\Sigma(r)=\Sigma_0 \left(\frac{r_0}{r}\right)\,\exp\left(-\left[\frac{r}{r_d}\right]^\eta\right)
\end{equation}
with $r_0=400\,\mathrm{au}$ and $\Sigma_0=1\,\mathrm{g/cm^2}$. Our standard
model has $r_d=150\,\mathrm{au}$ and $\eta=1$, leading to a disk gas mass of about
$M_{\mathrm{disk}}\simeq 0.039\,M_{\odot}$. This model is consistent with a
typical Lynden-Bell \& Pringle viscous disk model, with a power law
behavior close to the star and a weak exponential tail in the outer disk
regions.  But in Section \ref{sec-subkep} we will also show the results of
$\eta=2$, which matches the outer regions better, and hints toward an outer
truncation of the disk.

The full axisymmetric 2-D/3-D gas density structure is constructed from
$\Sigma(r)$ by demanding hydrostatic equilibrium in cylindrical coordinates
$(r_{\mathrm{cyl}},z)$. The procedure for this is described in Appendix
\ref{sec-twodee-structure}. This also uniquely determines the azimuthal velocity
$v_\phi(r_{\mathrm{cyl}},z)$. We choose a radial grid between $r_{\mathrm{cyl}}=10\,\mathrm{au}$ and
$r_{\mathrm{cyl}}=800\,\mathrm{au}$ with 128 grid cells logarithmically spaced. The vertical
extent of the $z$-grid is dependent on $r_{\mathrm{cyl}}$: From the midplane upward we set up
an equally-spaced $z$-grid of 64, with the top of the grid proportional to the
cylindrical radius: $z_{\mathrm{max}}(r_{\mathrm{cyl}})=0.7\,r_{\mathrm{cyl}}$. We assume mirror-symmetry
in the equatorial plane.

Once the model is set up in cylindrical coordinates, we map the model onto a
spherical coordinate grid $(r_{\mathrm{spher}},\theta)$, because the radiative
transfer tool we use, the {\small\tt RADMC-3D} package, requires spherical
coordinates.  We choose the same logarithmic gridding for $r_{\mathrm{spher}}$
as we did for $r_{\mathrm{cyl}}$. The $\theta$-grid has 64 grid cells linearly
spaced between $\pi/2-0.7\le\theta\le\pi/2$ (where $\theta=0$ is the north pole,
$\theta=\pi/2$ is the equator and $\theta=\pi$ is the south pole).

We choose a simple analytic prescription for the gas temperature, inspired from
the gas temperatures we actually measure with our method (see Section
\ref{sec-comaps-hd163296}, Eqs.~\ref{eq-midplane-temp} and
\ref{eq-surface-temp}). We could, instead, compute the temperature structure
self-consistently using Monte Carlo radiative transfer. However, the dust
optical depth at large radii is low. This means that, for the radii we are
interested in here ($r\gtrsim 50\,\mathrm{au}$), the ``warm surface, cool
midplane'' temperature structure is primarily dependent on the conical
midplane-shadow cast by the more dense inner disk regions ($r\ll
50\,\mathrm{au}$). Given that these inner regions are not the target of our
study, a prescribed temperature profile will do just as well. The prescription
is described in Appendix \ref{sec-temp-prescription}

In our model we will focus on the CO 2-1 line. In principle we may need to
worry about dust extinction, especially close to the star (inward of about
10 au) or in the known dust rings. In this model we will assume, however, that
the regions we probe are sufficiently dust-depleted that the dust extinction
does not play a role. 

The most uncertain parameter of our model is the $^{12}$C$^{16}$O abundance. We use an
abundance of $10^{-4}$ relative to H$_2$. We do not account for freeze-out or
photodissociation of CO.  For the case of HD 163296, the star is a Herbig star,
the disk midplane temperature is anyway likely to be above 20 K (if only
marginally) throughout the disk, meaning that freezeout is probably avoided.
For T Tauri stars this may be a different matter. Photodissociation is in
principle important for the CO abundance in the surface layers. It can be
included in a simplified way using a simple column density threshold, along the
lines of \citet{2013ApJ...774...16R}. But since the total CO column density of
the warm surface layers is not our real concern in this paper, we ignore a
treatment photodissociation. The CO molecular data are taken from the LAMDA
database in Leiden\footnote{\url{http://www.strw.leidenuniv.nl/~moldata}}. We focus on the $J=$ 2-1 line.  With {\small\tt RADMC-3D} we
can now compute channel maps of the CO 2-1 line. We assume that the level
populations of the CO molecules are in LTE. Microturbulence is not included.

\subsection{Resulting channel maps}
\begin{figure}
\includegraphics[width=0.5\textwidth]{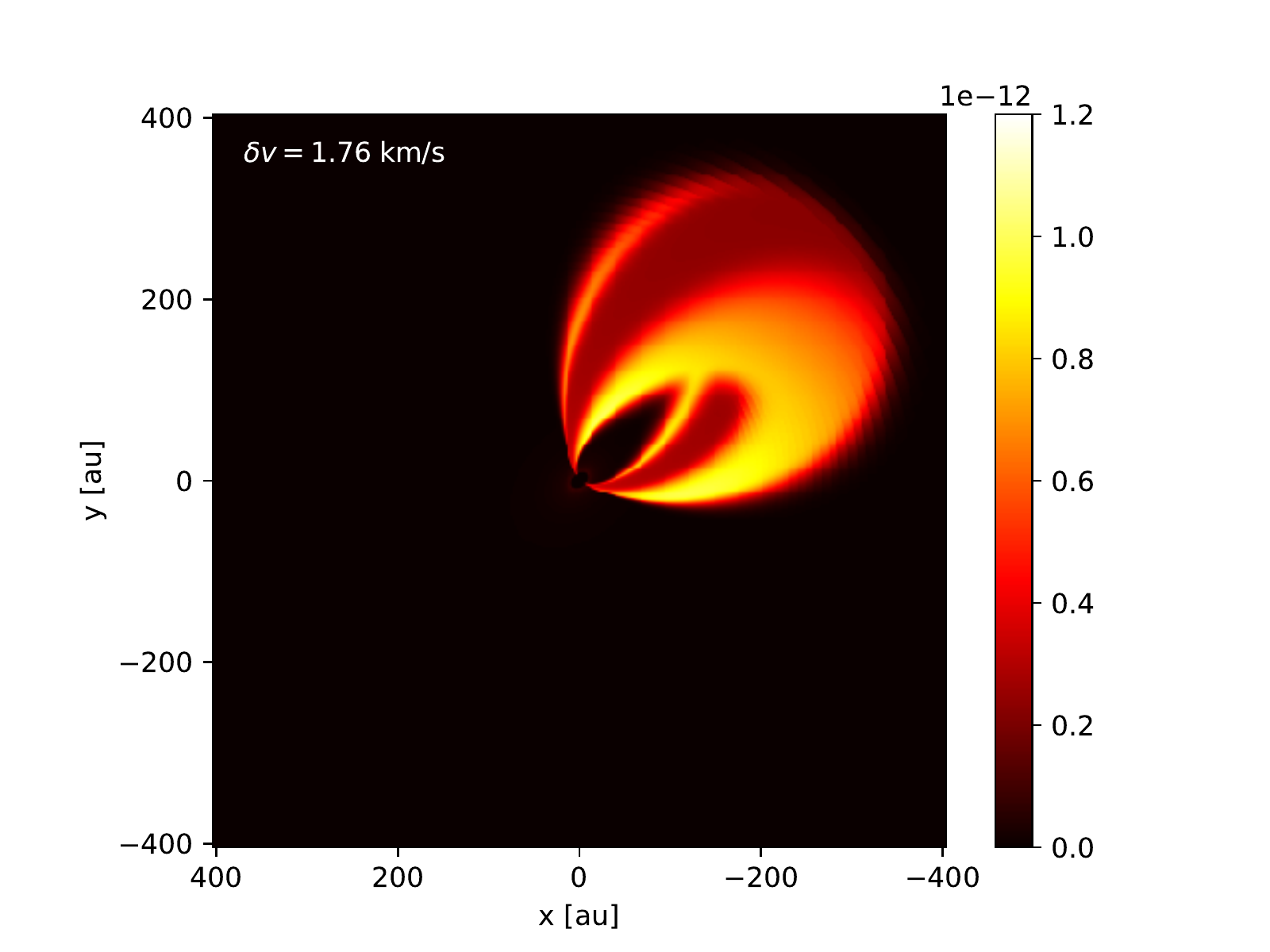}
\caption{\label{fig-ear-3d}The CO 2-1 channel map of the model of the disk of HD
  163296 at $\delta v=1.76\,\mathrm{km/s}$. The color scale stretches from $I_\nu=0$
  to $1.2\times 10^{-12}\,\mathrm{erg\,cm^{-2}\,s^{-1}\,Hz^{-1}\,ster^{-1}}$. The
  same color scale is used for all subsequent channel map plots.}
\end{figure}

In Figure \ref{fig-ear-3d} the channel map at $\delta v=1.76\,\mathrm{km/s}$ is
shown, where $\delta v$ is defined as the line-of-sight velocity relative to the
systemic velocity ($5.8\,\mathrm{km/s}$)  of the system. This image shows the familiar Keplerian
``teacup handle'' shape: a teardrop-shaped ring of emission from the locations
where the projected Kepler velocity is $1.76\,\mathrm{km/s}$ redshifted. In
fact, the ``teacup handle'' has a bright front and back layer, separated by a
dim region in between. These bright front and back rings, which look like
the twin wings of a dragonfly, come from the CO emission from the warm surface
layers \citep{2007ApJ...669.1262P}. In between the two dragonfly wings the
emission is much weaker, because the temperature in the disk midplane region is
much lower than in the surface layers.

If the CO emission were optically thin, then both the front-side and the
back-side ``dragonfly wings'' would be equally bright. And they would both be
closed teardrop shaped rings.  We would not be able to distinguish between front
side and back side. The dust opacity is negligible in this model at this
wavelength and distance from the star. For the real disk around HD 163296 the
dust extinction plays a role, but only near the known dust rings at 67 and 100
au \citep{2018ApJ...869L..49I}.

\begin{figure}
\includegraphics[width=0.5\textwidth]{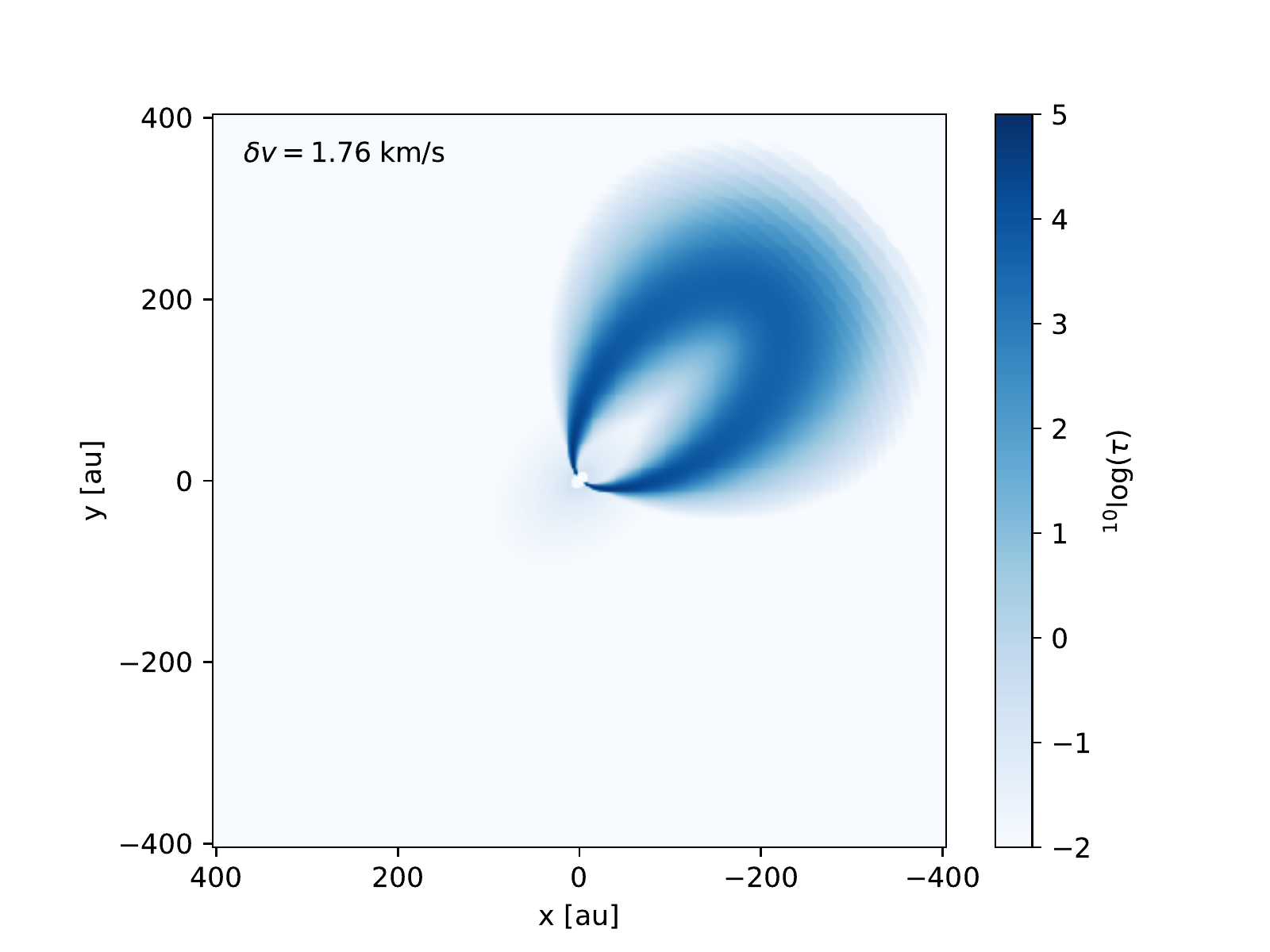}
\caption{\label{fig-ear-3d-tau}The optical depth of the CO 2-1 channel map
  shown in Fig.~\ref{fig-ear-3d}. The color scale is logarithmic, spanning from
  $\tau=10^{-2}$ to $\tau=10^{+5}$.}
\end{figure}

However, the CO line emission is highly optically thick
(see Fig.~\ref{fig-ear-3d-tau}). The space between the two
dragonfly wings is therefore not ``hollow'': it is highly opaque. And it emits
optically thick line emission at the midplane temperature.

Therefore the back side CO emission is extincted by the cooler CO gas at the
midplane. Only a narrow rim of the bright CO emission from the back side is
seen. And at some point this backside ``dragonfly wing'' disappears entirely
behind the optically thick cool CO emission. Only the front-side bright wing is
complete.

This geometry is a well-known outcome of radiative transfer models of line
emission channel maps of protoplanetary disks
\citep[e.g.][]{2007ApJ...669.1262P, 2013ApJ...774...16R, 2018ApJ...860L..13P}.

To better understand the geometry of what we see in Fig.~\ref{fig-ear-3d}, it is
useful to realize that the teacup handle shaped object (from lower to upper
surface, including the opaque medium between the dragonfly wings) is highly
optically thick, while the rest of the disk is transparent at this
wavelength. One can regard this teacup handle shaped object thus as a truly 3-D
object. The surface of this 3-D object is the $\tau\sim 1$ surface.  But this
surface is not only at the top and the bottom (the disk surface), but also at
the inner and outer edge of the ``teacup handle''. The locations of the inner
and outer edge depend on the velocity channel one is looking at, and the
intrinsic line width. At any given channel we are looking at the surface of the
``teacup handle''. The brightness is, with the Eddington-Barbier rule, the
Planck function at the temperature of the gas at this surface. The dark lanes
between the bright dragonfly wings represents the Planck function at the lower
temperature of the midplane. By looking at the inner edge and the outer edge of
the ``teacup handle'' we can thus directly measure the gas temperature inside
the disk, as a function of radius. Different channels will allow us to obtain
several independent ``cuts'' through the disk, each yielding the temperature as
a function of radius, though in a slightly different range in $r$.

\subsection{Extracting intensity along iso-velocity curves}
Let us test the Eddington-Barbier temperature extraction with our model. To
associate the projected Kepler velocity with the location on the channel map, we
need to compute the iso-velocity curves corresponding to the channel: the curves
for which the projected Kepler velocity along the line-of-sight equals $\delta
v$.


The azimuthal velocity $v_\phi(r)$ is the Kepler velocity,
i.e.\ $v_\phi(r)=v_K(r)=\sqrt{GM_{*}/r}$. The iso-velocity curve is defined by
the equation
\begin{equation}\label{eq-iso-velo}
v_\phi(r)\,\cos\phi = \frac{\delta v}{\sin i}
\end{equation}
which gives a relation between $r$ and $\phi$ in the plane of the disk. This
equation is also valid for the case when $v_\phi(r)$ deviates from Kepler (see
Appendix \ref{sec-twodee-structure}). And if we know how $v_\phi(r,z)$ varies
with height $z$ above the midplane, we can also use it to compute isovelocity
curves for line emission from above and below the midplane. For every pair
$r,\phi$ (and possibly $z$) we can now project this location onto the sky
\citep[see e.g.~the appendix of][]{2018ApJ...869L..49I}.

\begin{figure}
\includegraphics[width=0.5\textwidth]{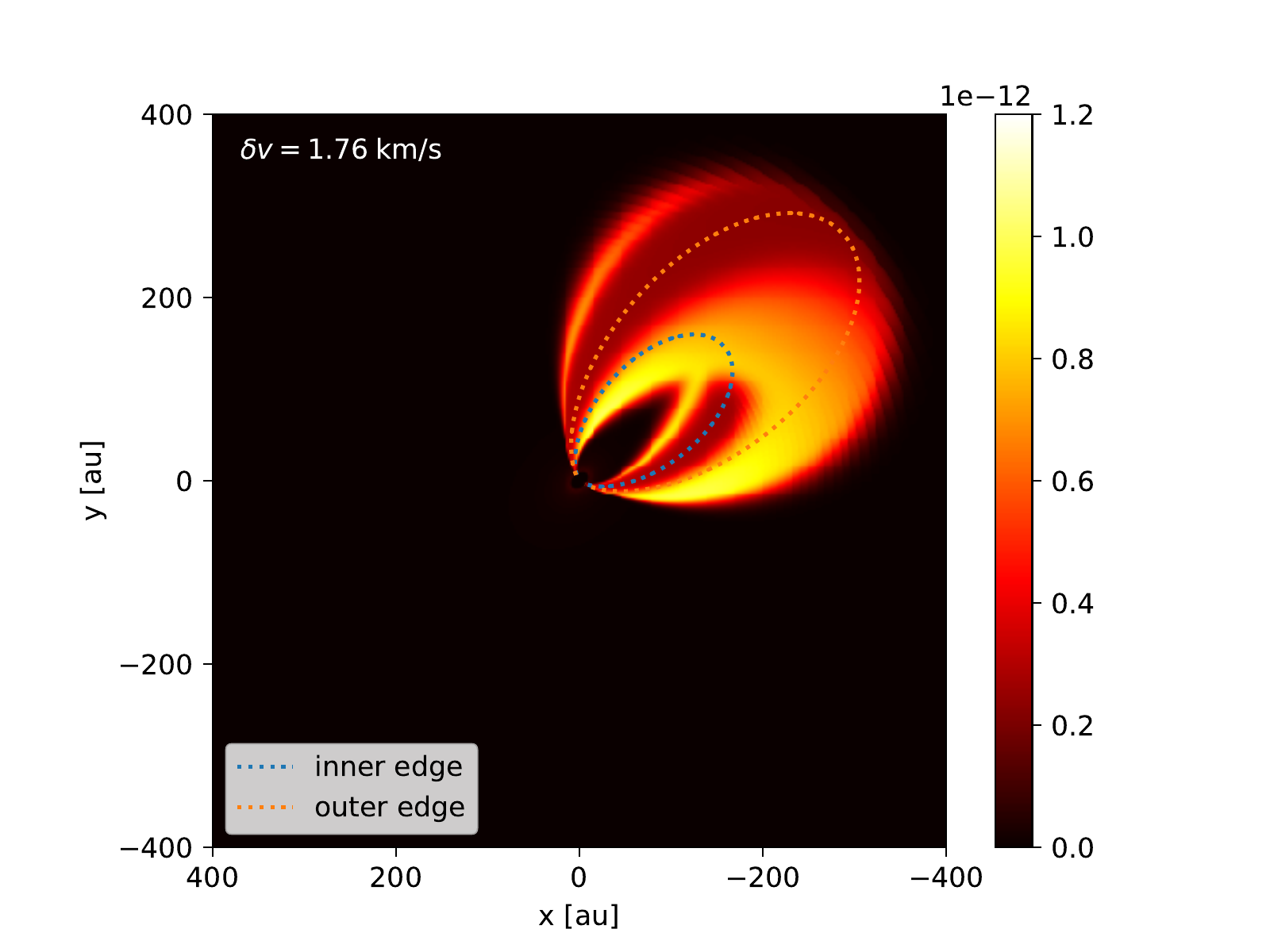}
\caption{\label{fig-ear-with-paths}Same as Fig.~\ref{fig-ear-3d}, but now with
  two paths overplotted that represent equal-radial-velocity curves in the
  midplane. The inner one is chosen to probe the inner edge of the 3-D ``teacup
  handle'', while the outer one is chosen to probe the outer edge.}
\end{figure}

In Fig.~\ref{fig-ear-with-paths} the same ``teacup handle'' is shown as before, but
now we overplot the midplane iso-velocity curves for two different velocities. The inner
curve is for a $0.35\,\mathrm{km/s}$ higher velocity, while the outer curve is
for a $0.20\,\mathrm{km/s}$ lower velocity than the $\delta
v=1.5\,\mathrm{km/s}$.  These roughly probe the inner and outer surface of the
``teacup handle''.

\begin{figure*}
\centerline{\includegraphics[width=0.5\textwidth]{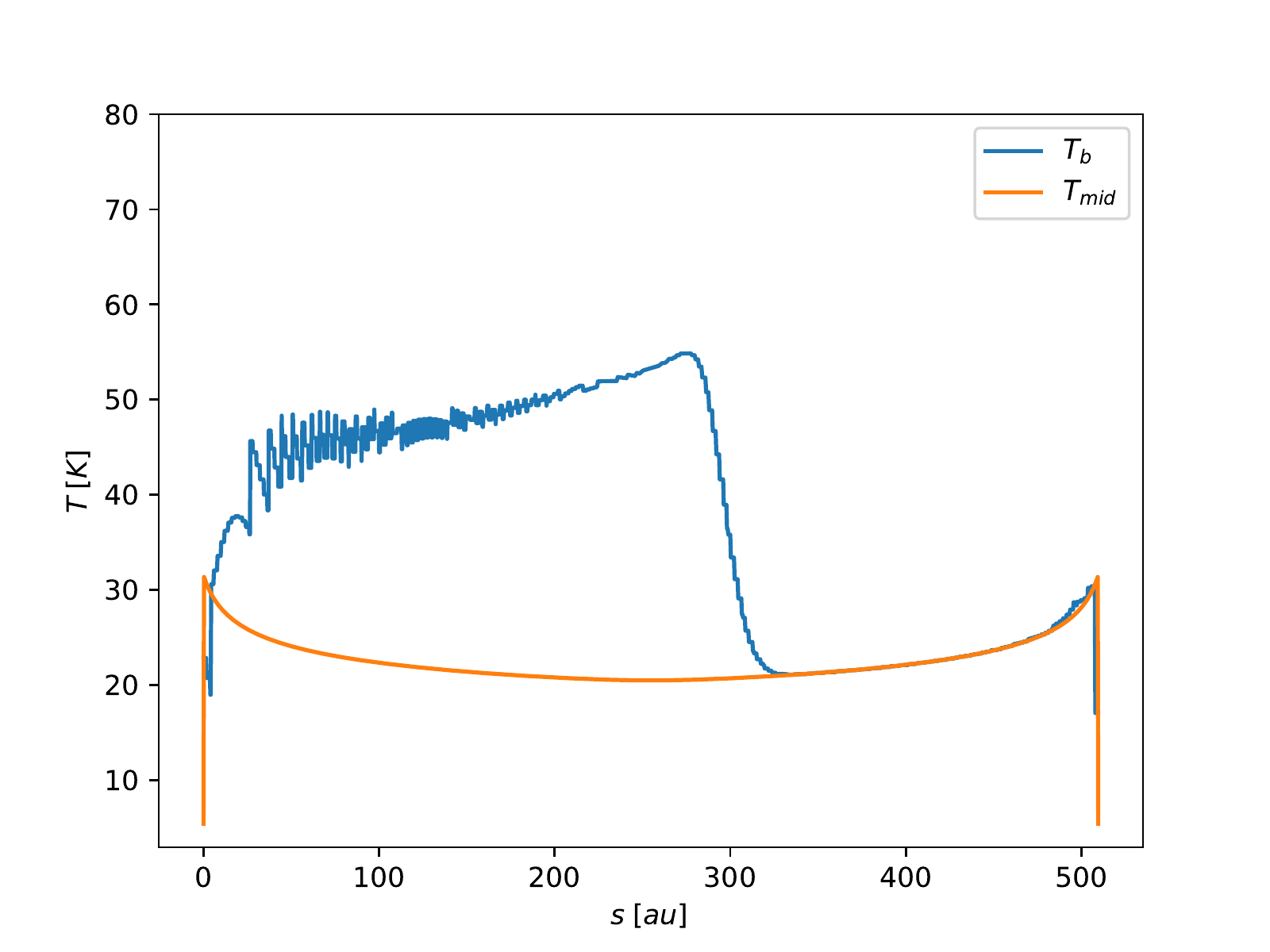}
\includegraphics[width=0.5\textwidth]{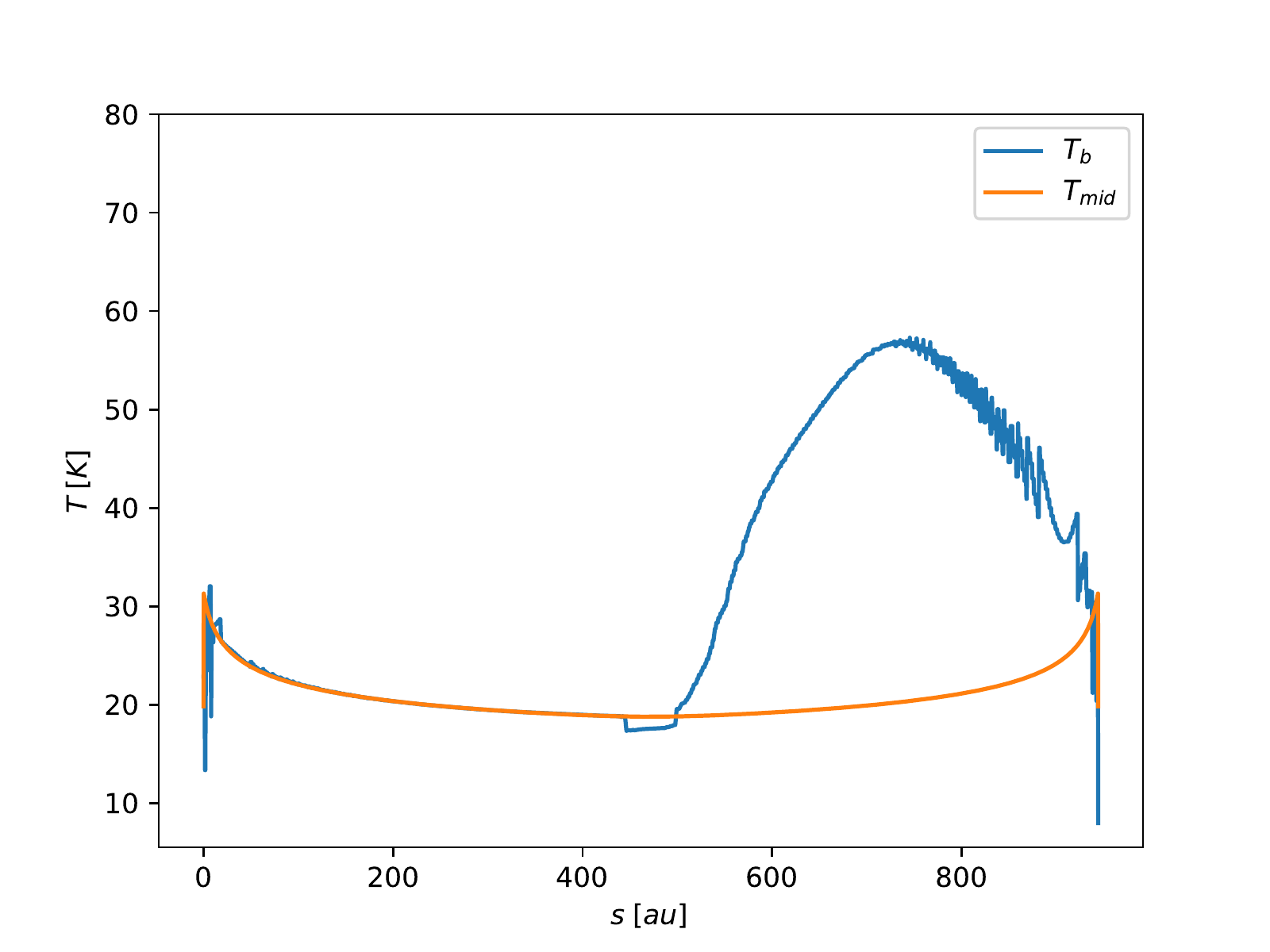}}
\caption{\label{fig-paths}The brightness temperature (using the full Planck
  function) along the two paths shown in Fig.~\ref{fig-ear-with-paths}. Left:
  the inner path. Right: the outer path. Overplotted is the actual midplane
  temperature of the model. In the regions where the path passes through the
  part of the image where the darker midplane emission is visible, the match
  between the synthetic observation and the true gas temperature is nearly
  perfect, as expected. The deviation between the two curves elsewhere is
  due to the fact that the path passes through the much brighter surface layer
  emission.}
\end{figure*}

In Fig.~\ref{fig-paths} the intensity profiles along these paths are shown.  The
intensity is represented in the form of brightness temperature $T_b$ in units of
Kelvin. The measured intensity $I_\nu$ and the brightness temperature $T_b$ are
related via the Planck function according to
\begin{equation}\label{eq-tb-vs-inu}
I_\nu \equiv B_\nu(T_b)
\end{equation}
The Eddington-Barbier rule simply states that the measured brightness
temperature $T_b$ is about equal to the actual gas temperature
$T_{\mathrm{gas}}$ at the location along the line-of-sight where the disk
becomes optically thick. If this location is in the disk midplane, then we write
$T_{\mathrm{gas}}$ as $T_{\mathrm{mid}}$. Overplotted in that figure is the
actual midplane temperature. The paths start near the star, and then follow the
paths in clockwise direction.

One can see that, as expected, the curves of $T_b$ and $T_{\mathrm{mid}}$ match
very well in the region where the paths probe the inner or outer edge of the
``teacup handle'', which represent the disk midplane. Indeed the Eddington-Barbier
rule works well, and we can thus use it to probe directly the midplane
temperature.  In contrast, where the path passes along the bright emission from
the front surface layer, we instead get a much brighter intensity, because,
again according to Eddington-Barbier, we probe the surface layers.

In reality we measure the channel maps as an average over a certain channel
width $w$, i.e.\ averaged between $\delta v-w/2$ and $\delta v+w/2$, instead of
at the precise velocity $\delta v$. We investigate this in Appendix
\ref{app-channel-width}. It turns out that the effect on the measured midplane
temperature is relatively small for channel widths up to a few times the
intrinsic line width.

One should also worry about beam smearing and noise, both of which limit the
feasibility of the method close to the star. And, finally, the effect of
turbulent broadening may have an effect similar to the channel-averaging: it will
widen the surface ``wings'' a bit more than the thermal broadening does (see
Appendix \ref{app-turbulence}). In fact, this ``wing-widening'' is what makes
it, in theory, possible to measure the microturbulence in the disk. For HD
163296, however, only an upper limit could be determined
\citep{2015ApJ...813...99F}.

Nevertheless, with sufficient resolution and sensitivity,
the method we present here constitutes a powerful method to directly measure the
midplane temperature of a protoplanetary disk, without having to make any model
assumptions.

\section{Analysis of the observed CO channel maps of HD 163296}
\label{sec-comaps-hd163296}
Armed with the knowledge of Section \ref{sec-radtrans-model} we now turn to the
CO $J=$2-1 channel maps of the disk around HD 163296 of the DSHARP campaign
\citep{2018ApJ...869L..41A, 2018ApJ...869L..49I}, with the goal of directly
measuring the midplane temperature of this disk. The source HD 163296 is one of
the best studied protoplanetary disks \citep[e.g.][]{2000ApJ...544..895G,
  2012A&A...538A..20T, 2013ApJ...774...16R, 2016PhRvL.117y1101I,
  2018ApJ...860L..13P, 2018A&A...614A..24M, 2018ApJ...869L..49I,
  2018ApJ...857...87L, 2019ApJ...882L..31B, 2019Natur.574..378T}. 

\begin{figure*}
\centerline{\includegraphics[width=0.95\textwidth]{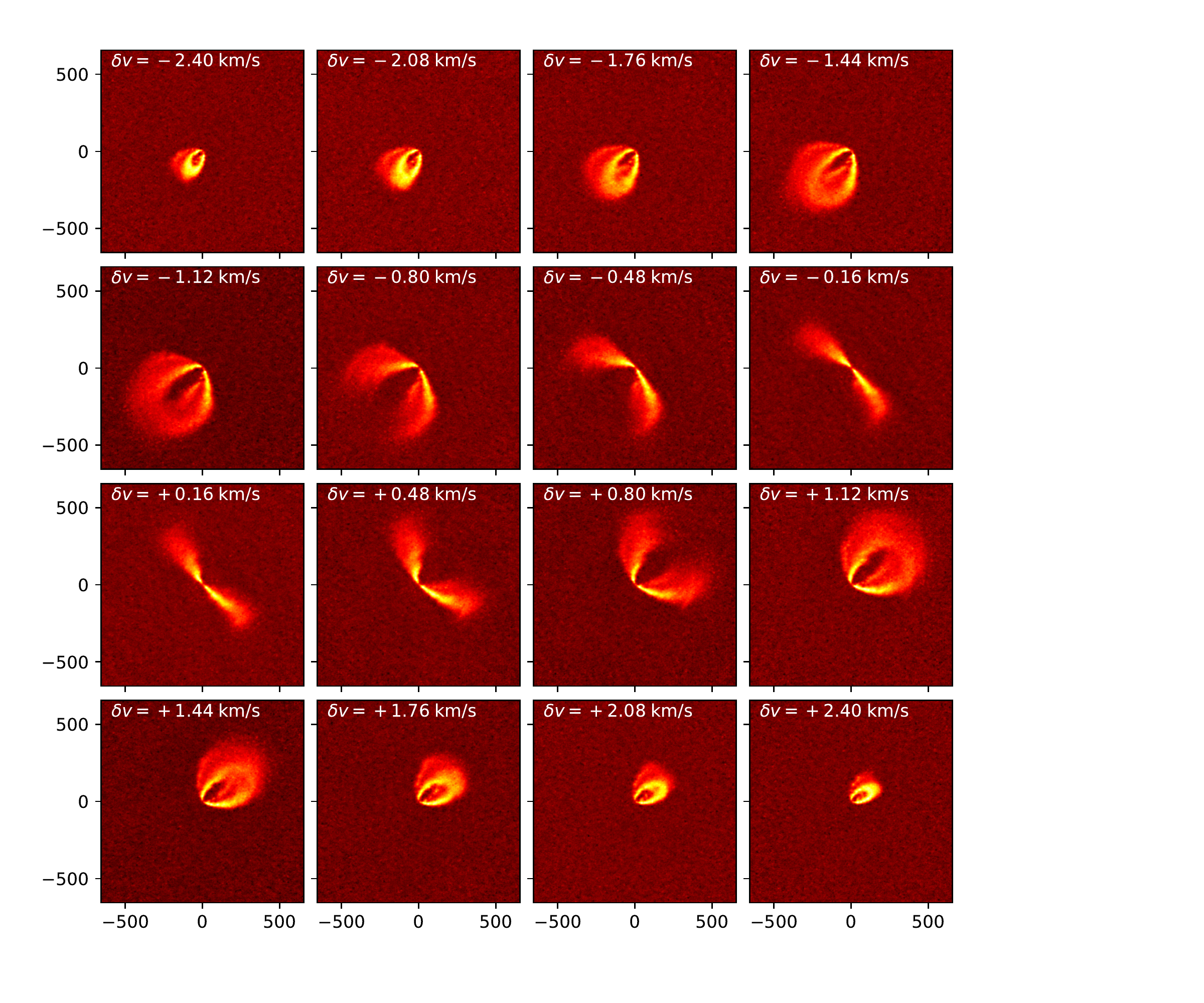}}
\caption{\label{fig-all-channels}Observed channel maps of CO $J=$2-1 of HD
  163296. The channel velocity $\delta v$ is relative to the system velocity.
  The horizontal and vertical coordinates are in au.}
\end{figure*}

Fig.~\ref{fig-all-channels} shows the channel maps, the top eight of which show
the blue-shifted channels relative to the systemic velocity, while the bottom
eight show the red-shifted channels.  Note that the extent of the maps ($\sim
650\,$au), and therefore of the CO emission, is much larger than the radius of
the outermost massive dust ring at 100 au
\citep{2018ApJ...869L..49I,2018ApJ...869L..42H}.

In using the Eddington-Barbier rule, we assume that the CO lines of this disk
are optically thick and in LTE throughout the vertical structure of the disk,
all the way out to about 400 au. The models of Section \ref{sec-radtrans-model}
show that without any freeze-out, the optical depth of the CO lines at line
center can easily be of the order of 1000 or more. Partial freeze-out of CO is
therefore possible without violating our assumption of the line being optically
thick. For instance, \citet{2015ApJ...813..128Q} find that CO freeze-out reduces
the gas-phase CO abundance by a factor of about 10, which would in our case
easily keep the optical depth of the CO well above unity. \revised{But freeze-out
  is nonetheless an important caveat, which will be discussed in more detail
in Section \ref{sec-freezeout}.}

We also assume that the dust is optically thin everywhere. This is not
correct around 100 au and 75 au, the locations of known dust rings. But the
method of extraction of the midplane gas temperature described in Section
\ref{sec-radtrans-model} requires the ``teacup handle'' to be well-resolved, which
limits us to regions beyond 100 au anyway. 

From the channel maps of Fig.~\ref{fig-all-channels} we see that the ``teacup
handle'' of the channels between $\delta v=-0.8\,\mathrm{km/s}$ and $\delta
v=+0.8\,\mathrm{km/s}$ are not closed. The CO emission appears to end somewhere
between 400 and 500 au, cutting off the outer parts of the ``teacup
handle''. We will omit these channels from our analysis from here onward,
even though our technique can, to a certain extent, also be applied to
these channels.

The remaining channels all show very clear teacup handle shapes. All of
these images show the expected optical depth effects discussed in Section
\ref{sec-radtrans-model}. The bright emission is from the warm disk surface
facing us. The emission from the warm disk surface on the other side is mostly
extincted, with the exception of a narrow lining, also seen in the radiative
transfer models. \revised{For the positive velocity channels (and mirrored for the
  negative velocity channels),}
this narrow lining is only visible on the south-west side for
the inner boundary of the ``teacup handle'' and to the north-east side for the
outer boundary, exactly as predicted by the radiative transfer models. The
disappearance of the lining on the north-east for the inner boundary and
south-west for the outer boundary of the ``teacup handle'' is due to the CO
self-absorption. It is therefore clear that there is cooler gas between the two
warm surface layers, which is rich enough in CO to be optically thick in the
line. The simplest explanation is that the CO at the disk midplane is not frozen
out entirely, or not at all. The midplane is thus optically thick in the CO
line, meaning that the Eddington-Barbier method for measuring the midplane
temperature should work. A slightly more complex scenario could be that the CO
is frozen out near the midplane, but not in the layers just below the warm
surface layers \citep[see e.g.\ the illustration in Fig.~3
  of][]{2018A&A...609A..47P}. In that case, our method would not measure the
midplane temperature, but just the temperature slightly above the midplane.
\revised{We will discuss this scenario in Section \ref{sec-freezeout}.}

\subsection{Midplane temperature}
\label{sec-midplane-temperature}
\begin{figure*}
\centerline{\includegraphics[width=0.95\textwidth]{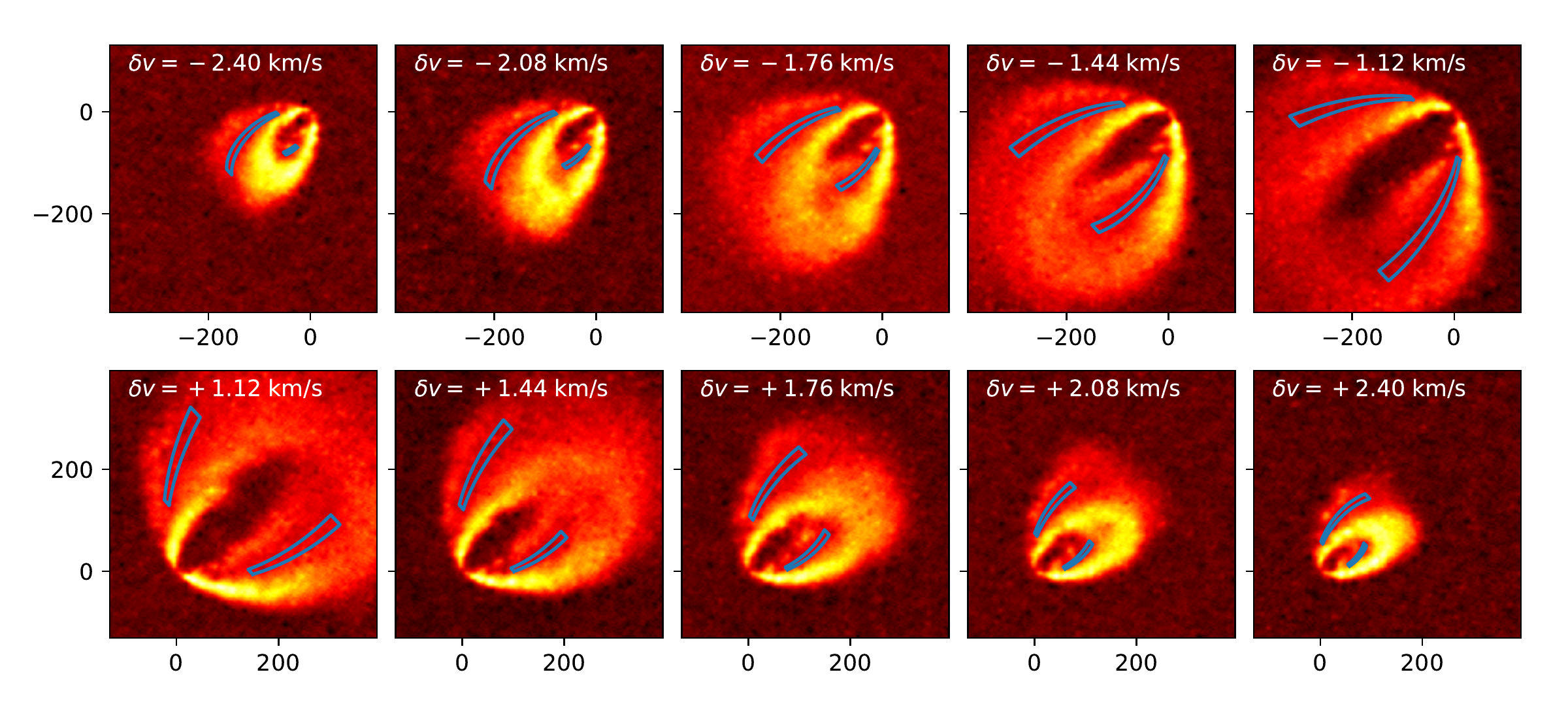}}
\caption{\label{fig-obs-chanmap-mid}Observed channel maps of CO $J=$2-1 of HD
  163296.  The channel velocity $\delta v$ is relative to the system velocity.
  Overplotted at the areas where the midplane temperature is measured. In each
  panel the top area probes the outer edge of the ``teacup handle'', while the
  bottom area probes the inner edge.}
\end{figure*}
To probe the midplane temperature we will measure the intensity $I_\nu$ at
locations in the channel maps that we estimate to be the inner- and outer edges
of the ``teacup handle'' at the midplane. The regions at which we probe $I_\nu$
for this purpose are indicated in the channel maps in
Fig.~\ref{fig-obs-chanmap-mid}.

These regions (two per panel) were constructed using iso-velocity curves. Given
that a protoplanetary disk does not rotate exactly at Keplerian rotation rate
(see Appendix \ref{sec-twodee-structure}), we slightly offset the iso-velocity
curves in velocity so that they nicely probe the inner and outer edge of the
``teacup handle'', as shown in Fig.~\ref{fig-obs-chanmap-mid}.

Given that each isovelocity curve should disappear behind the CO self-absorption
for more than half of its length (because geometrically, we can only see part of
the inner and outer edge of the ``teacup handle'' if it is optically thick), we
limit the curves accordingly. The exact choice of the limited domain along the
isovelocity curves is done by eye.  Given that the dim emission between the
``dragonfly wings'' becomes unresolved when coming too close to the star, we set
a minimum radius for each curve, which we also determine by eye. This procedure
yields two arcs for each panel along which to determine the brightness
temperature: one for the inner edge of the ``teacup handle'' and one for the outer
edge. For every point along these arcs, the corresponding radius is known
through the Keplerian velocity model. Finally, to gain signal-to-noise we puff
up each arc in vertical direction (with respect to the disk geometry) by
averaging over 10 arcs with aspect ratios ($z/r$) between -0.05 and +0.05. The
complete procedure described above defines not just two arcs but two areas in
the channel maps, as shown in Fig.~\ref{fig-obs-chanmap-mid}.

\begin{figure}
\centerline{\includegraphics[width=0.48\textwidth]{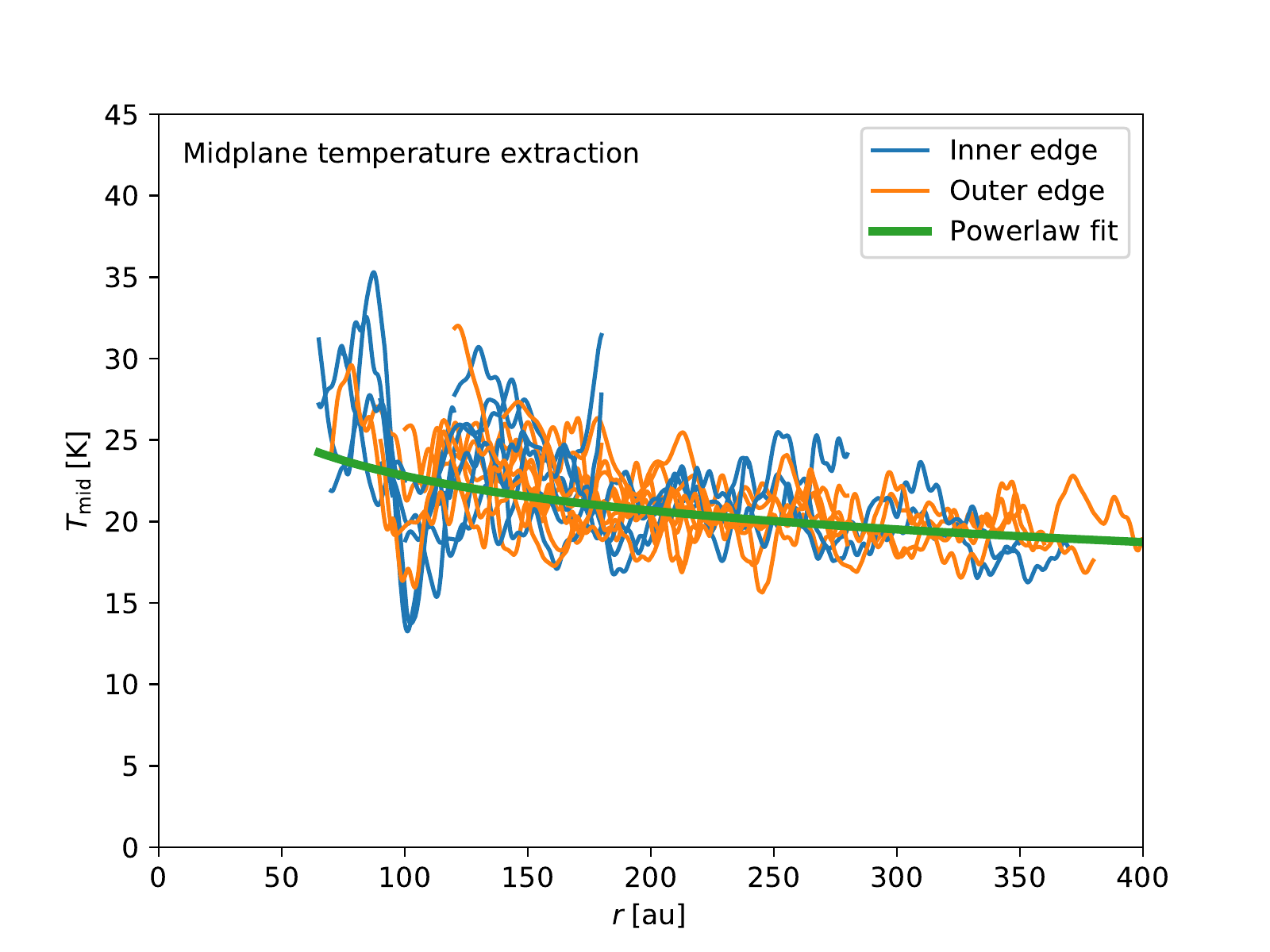}}
\caption{\label{fig-obs-temp-mid}Midplane temperature measurements within the
  various areas shown in Fig.~\ref{fig-obs-chanmap-mid}. The color blue is for the
  area on the inside of each ``teacup handle'', while the orange color is for
  the one on
  the outside. The green thick line is the best-fit parameterized model.}
\end{figure}

Each arc yields an intensity function $I_\nu(r)$, which is converted, using the
Planck function, into a brightness temperature $T_b(r)$
(Eq.~\ref{eq-tb-vs-inu}). The Eddington-Barbier rule says that this brightness
temperature is the actual gas temperature at the inner-/outer edge of the
``teacup handle''. Each arc is thus an independent measurement of
$T_{\mathrm{mid}}(r)$, but along varying segments of $r$. The results are
plotted in Fig.~\ref{fig-obs-temp-mid}.

The midplane gas temperature as a function of radius, as seen in this figure, is
almost constant, hovering around 20 K. It is a bit higher at 100 au (about 25 K)
and a bit lower at 400 au (about 18 K), but otherwise has a surprisingly shallow
slope. The noise is mostly around $\pm 5\,\mathrm{K}$, though inward of 200 au
it increases to about $\pm 10\,\mathrm{K}$. The temperature estimates from the
outer edges appear to be a bit less noisy than those from the inner edge. This
is presumably because the outer edge is better visible than the inner edge, and
there is thus less danger of accidently picking up flux from the bright surface
layers.

Using the least-squares method we fit a simple parameterized model to the data:
\begin{equation}\label{eq-midplane-temp}
T_{\mathrm{mid}}(r)\simeq 18.7\,\left(\frac{r}{400\,\mathrm{au}}\right)^{-0.14}\;\mathrm{K}
\end{equation}
which is shown in green in Fig.~\ref{fig-obs-temp-mid}.

\begin{figure}
\centerline{\includegraphics[width=0.48\textwidth]{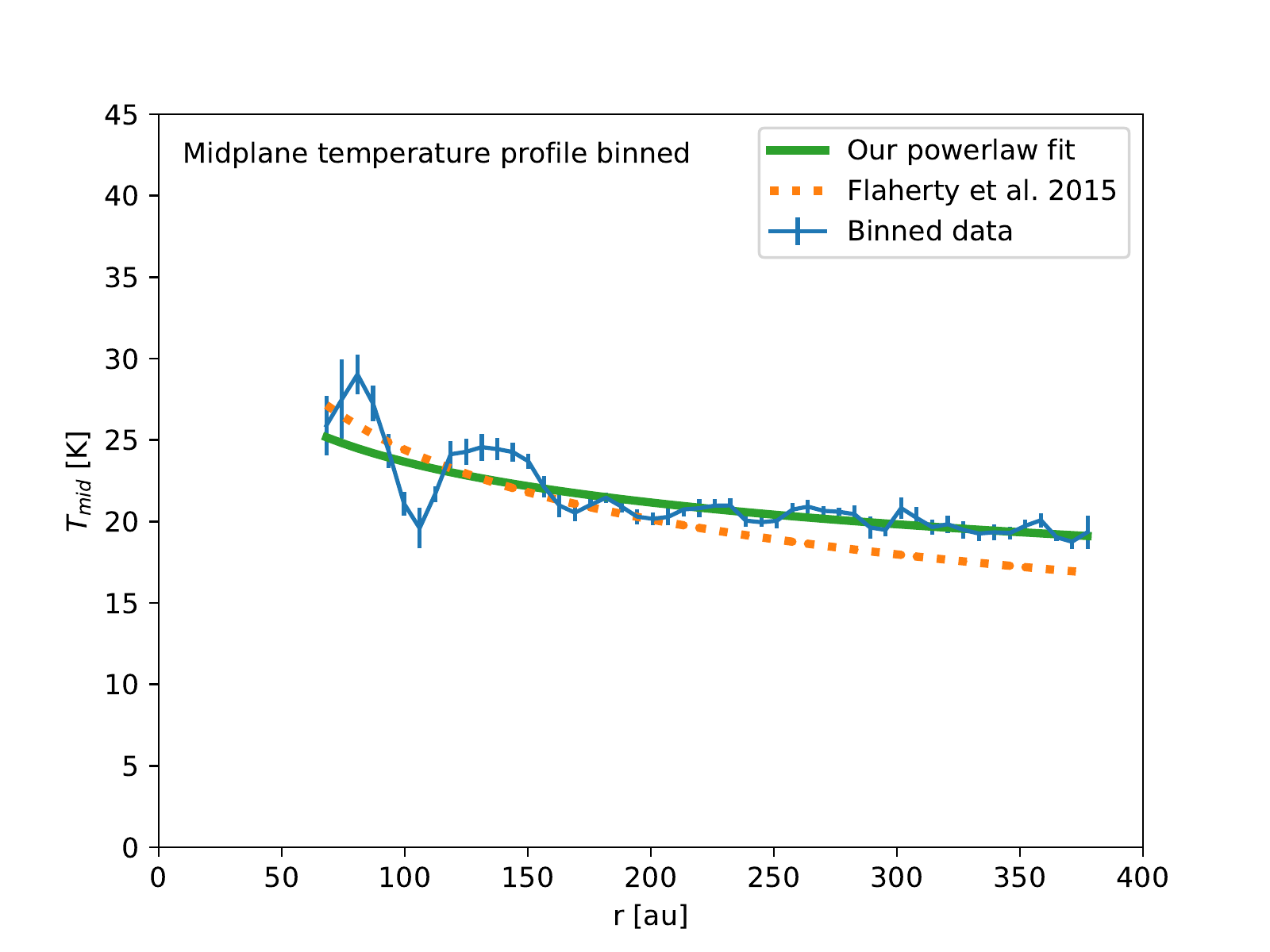}}
\caption{\label{fig-obs-temp-mid-bins-err}Midplane temperature measurements. As
  Fig.~\ref{fig-obs-temp-mid}, but now binned in radial bins.}
\end{figure}

Another way to analyze these $T_{\mathrm{mid}}$ data is by binning them in
radial intervals with the width of the average deprojected beam
$b_{\mathrm{avdeproj}}\simeq \sqrt{b_{\mathrm{maj}}b_{\mathrm{min}}}/\cos i$,
where $b_{\mathrm{maj}}$ and $b_{\mathrm{min}}$ are the FWHM beam sizes along
the major and minor axis of the beam respectively, and $i=46.7^{\circ{}}$ is the
inclination of the disk \citep[see Appendix H of][]{2018ApJ...869L..46D}.  For
each radial bin we figure out which of the tracks shown in
Fig.~\ref{fig-obs-chanmap-mid} contributes to this bin. Each track, when binned,
thus yields an independent measurement of the temperature in that bin. It is not
easy to estimate the error on each of these measurements. Instead, we compute
the mean $\mu$ and the variance $\sigma^2$ of the measurements within the same
bin, and estimate the error as Err$\simeq \sigma/\sqrt{N}$ where $N$ is the
number of tracks contributing to that radial bin. The results are shown in
Fig.~\ref{fig-obs-temp-mid-bins-err}. 

The results show a wiggly curve. It is not clear if these wiggles are real or
just an artefact of the analysis method. One dip, however, stands out strongly:
the dip near 100 au. This is the result of the dust ring. It is a
combination of the extinction effect described in \citet{2018ApJ...869L..49I}
and the continuum-subtraction performed on the CO data.

Overplotted with a dotted line is the power law fit obtained by
\citet{2015ApJ...813...99F} for their high-resolution CO 3-2 data. Their fit is
$T_{\mathrm{mid}}(r)=21.8\,(r/150\,\mathrm{au})^{-0.278}\,\mathrm{K}$, which
they find using a forward radiative transfer modeling approach in combination
with an MCMC fitting procedure. Their finding is in good agreement with our
model-independent temperature measurement.

\subsection{Surface temperature}
In principle we can apply a very similar procedure to probe the surface
temperature. For each channel we just need to know the exact shape of the
front-side surface of the ``teacup handle'', and then sample the intensity along that
curve. However, given that this bright ring is rather narrow, a model of its
shape must be rather accurate to avoid accidently missing the bright
intensity. The shapes of these isovelocity curves on the surface of the disk
depend on the flaring/non-flaring geometry of the surface as well as on the
possible deviation from Keplerian rotation (see Appendix
\ref{sec-twodee-structure}). While these parameters can be extracted from the
observations, the procedure will be rather sensitive to the proper determination
of these parameters.

Earlier papers have determined the surface temperature profile in a different,
and more robust way. \citet{2018ApJ...869L..49I} derived, from the same
dataset, the following surface temperature profile:
\begin{equation}\label{eq-surface-temp}
  T_{\mathrm{surf}}(r)\simeq 87 - 56\,\left(\frac{r}{400\,\mathrm{au}}\right)
\end{equation}

\section{Subkepler rotation due to disk outer edge}
\label{sec-subkep}
Serendipitously, our analysis of the channel maps reveals tentative evidence for
subkepler rotation in the very outer regions of the disk around HD 163296. Such
a subkepler rotation can be produced by a rather sharp exponential cut-off of
the outer disk (a disk outer edge), leading to a strong negative pressure
gradient $d\ln p/d\ln r\ll 0$, which then causes the subkepler rotation. This
seems to be consistent with the fading away of the CO emission beyond 500 au as
seen in the low-velocity channel maps and in zero-moment maps.

\begin{figure*}
\centerline{\includegraphics[width=0.95\textwidth]{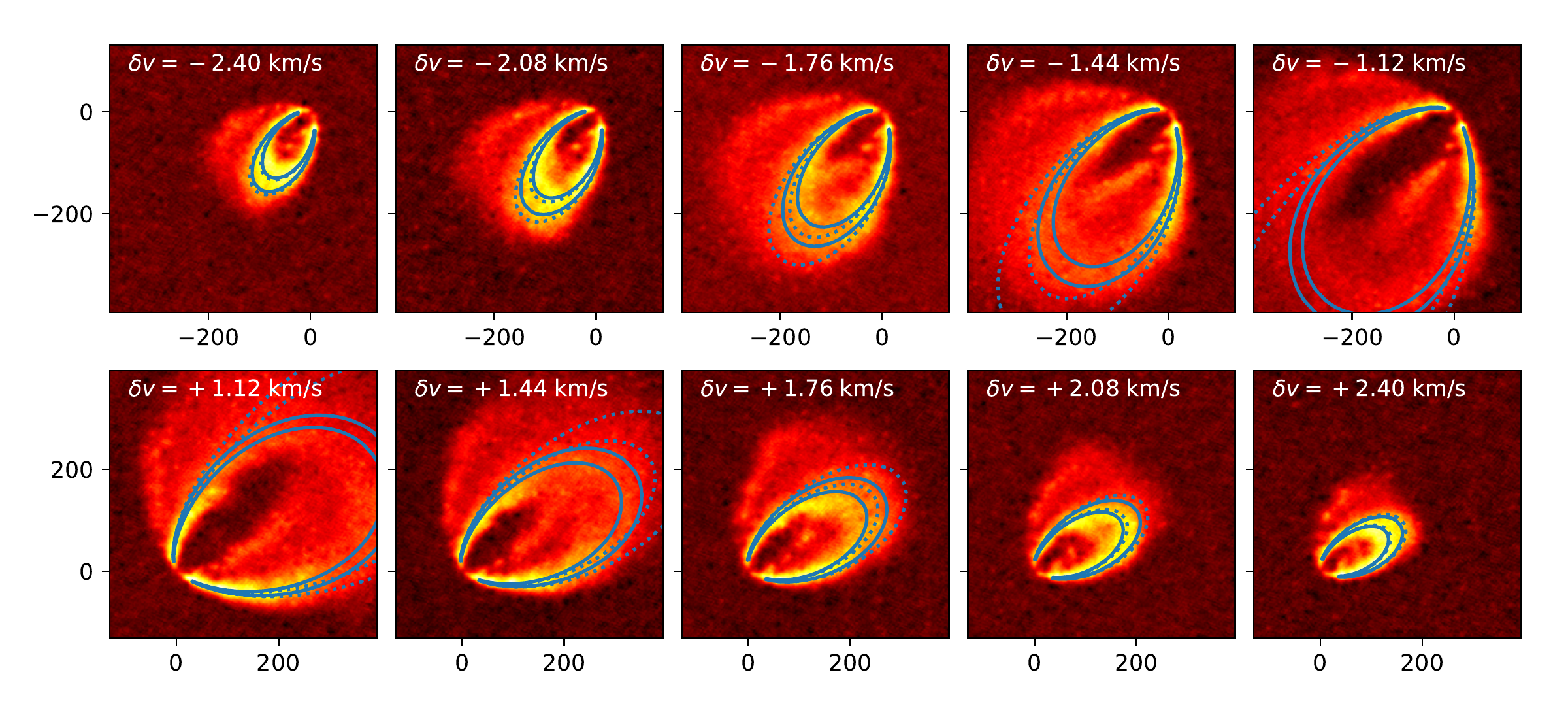}}
\caption{\label{fig-channels_with_kepcurves}As Fig.~\ref{fig-all-channels}, but
  now with the predicted locations of the CO surface emission for Keplerian disk
  (dotted line; corresponding to $\eta=1$) and subkeplerian disk (solid line;
  corresponding to $\eta=2$).}
\end{figure*}

The most direct way of spotting the evidence for the subkepler rotation is by
overplotting the expected shape of the ``surface wings'' of the ``teacup
handle'', assuming a disk without an outer edge (i.e. our standard model
given by Eq.~\ref{eq-surf-dens-model} with $\eta=1$). In that case the rotation
velocity of the gas is very close to Keplerian. This expected shape is seen as
the blue dotted lines in Fig.~\ref{fig-channels_with_kepcurves}. One can see
that the blue dotted lines predict the location of the surface emission rather
well for the high-velocity channels. But for the lower-velocity channels
($\delta v=\pm 1.12\,\mathrm{km/s}$ and $\delta v=\pm 1.44\,\mathrm{km/s}$)
the predicted ``surface wings'' are much larger than the data.

\begin{figure}
\includegraphics[width=0.5\textwidth]{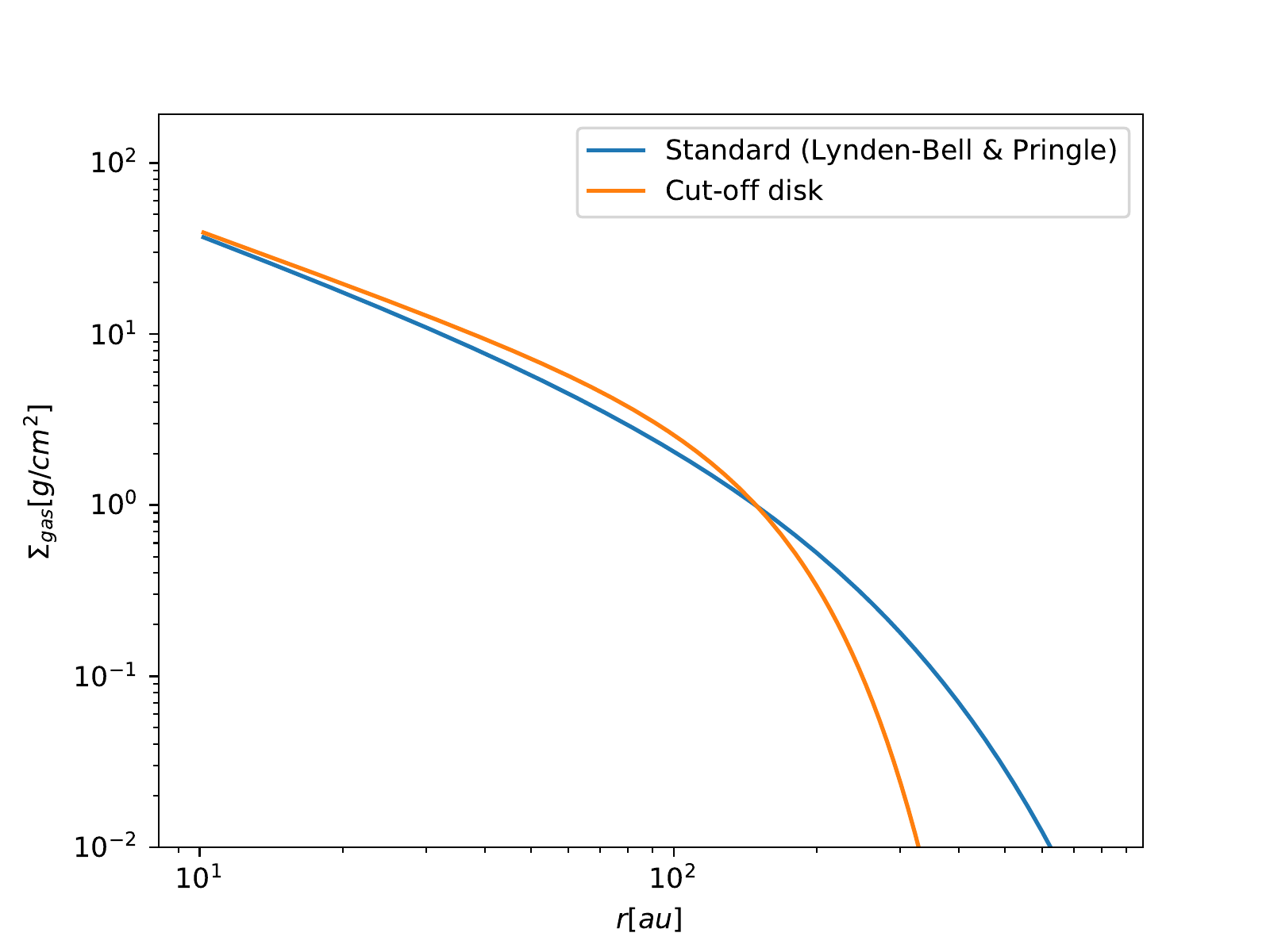}
\caption{\label{fig-sigma-gas-differences}Comparison of the surface density
  profile of the standard model of this paper ($\eta=1$, blue curve) and that of
  the disk with an outer edge ($\eta=2$, orange curve). For clarity, the
  axes are logarithmic.}
\end{figure}

\begin{figure}
\includegraphics[width=0.5\textwidth]{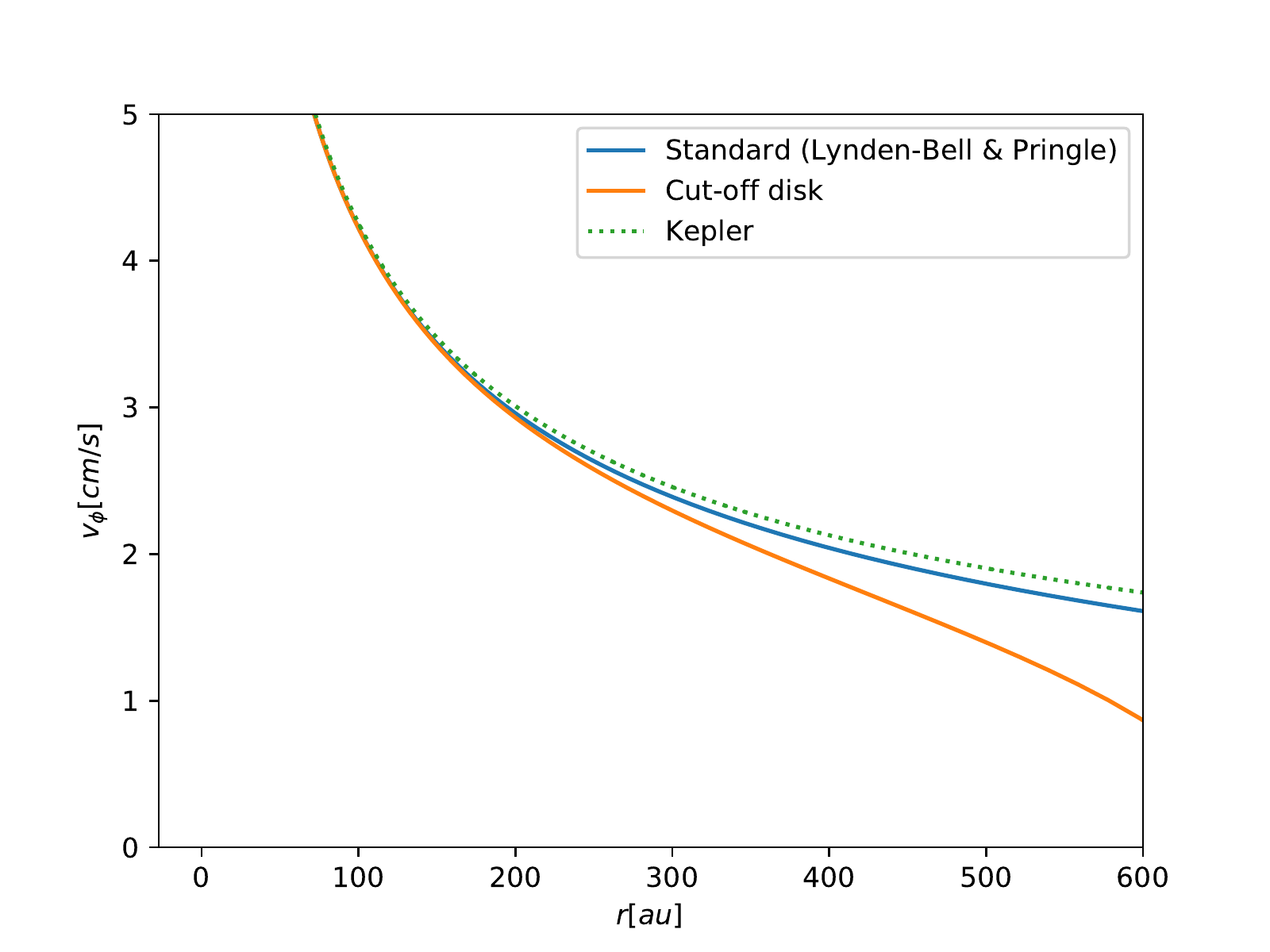}
\caption{\label{fig-vphi-gas-differences}Comparison of the midplane azimuthal
  gas velocity profile of the standard model of this paper ($\eta=1$, blue
  curve) and that of the disk with an outer edge ($\eta=2$, orange curve).
  Overplotted with a dotted line is the midplane Kepler velocity. In
  contrast to Fig.~\ref{fig-sigma-gas-differences}, the axes are here
  linear.}
\end{figure}

If we now modify our model to $\eta=2$, leading to a much stronger decay of the
surface density in the outer regions (see Fig.~\ref{fig-sigma-gas-differences}),
the pressure gradient in the outer disk regions becomes substantial, leading to
substantial deviation from Keplerian rotation (see
Fig.~\ref{fig-vphi-gas-differences}). The corresponding predicted emission
regions are shown with the blue solid lines in
\ref{fig-channels_with_kepcurves}. One can see that now the low-velocity
channels are much better matched.

\begin{figure*}
\centerline{\includegraphics[width=0.95\textwidth]{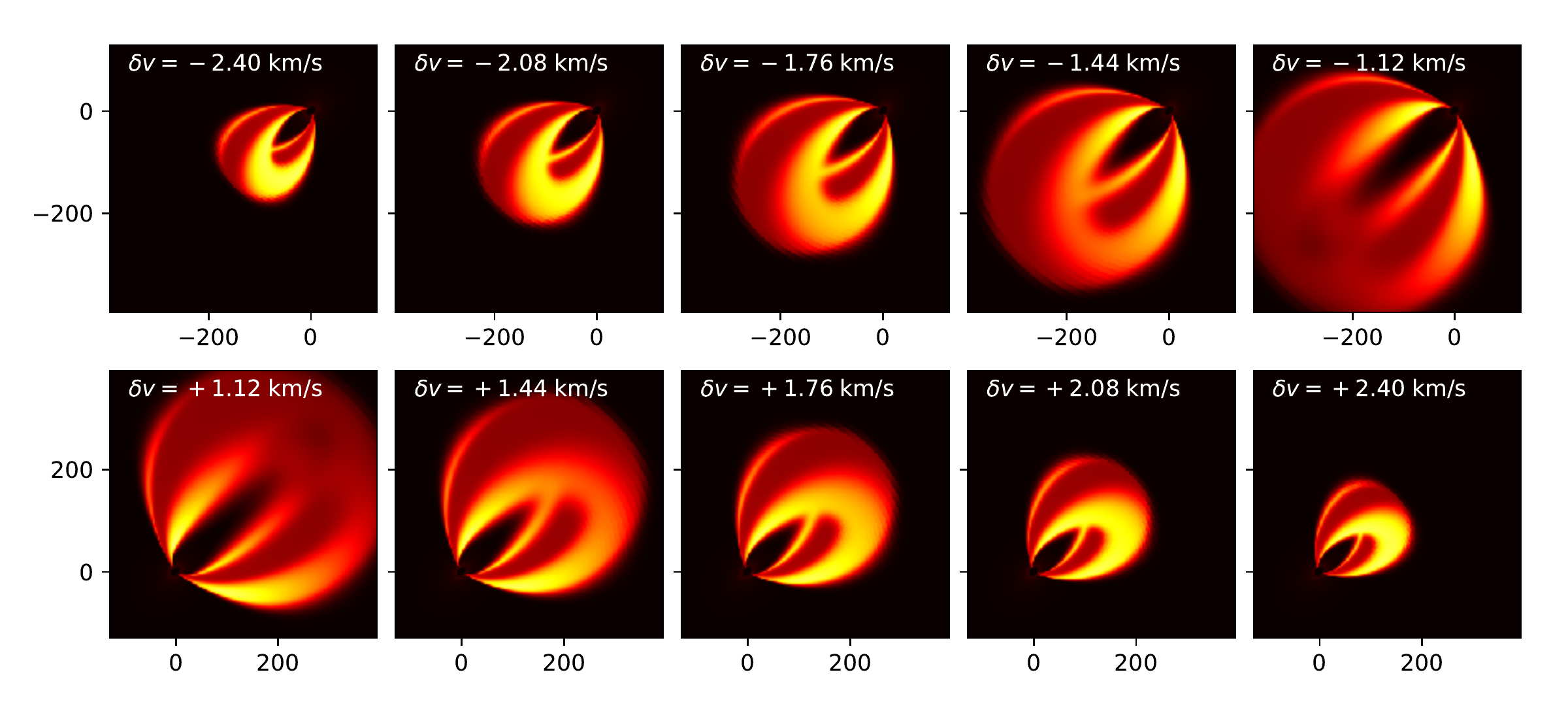}}
\caption{\label{fig-model-channels-yes-cutoff}
  Like Fig.~\ref{fig-obs-chanmap-mid}, but now the synthetic channel maps
  obtained from our {\small\tt RADMC-3D} model. Shown here is the case of $\eta=2$ (see
  Eq.~\ref{eq-surf-dens-model}), i.e.\ the model with an outer edge and,
  as a result, strong subkeplerian rotation.}
\end{figure*}

To confirm this, we need to compute the synthetic channel maps from the {\small\tt RADMC-3D}
radiative transfer model. In Fig.~\ref{fig-model-channels-yes-cutoff} one can
see the model predictions for the case of $\eta=2$ (all the rest of the parameters
are kept the same as in the standard model). In our view the match to the data
is reasonably good, even for the low-velocity channels, confirming that the
disk with a strong exponential cut-off ($\eta=2$) is consistent with the data.

\begin{figure*}
\centerline{\includegraphics[width=0.95\textwidth]{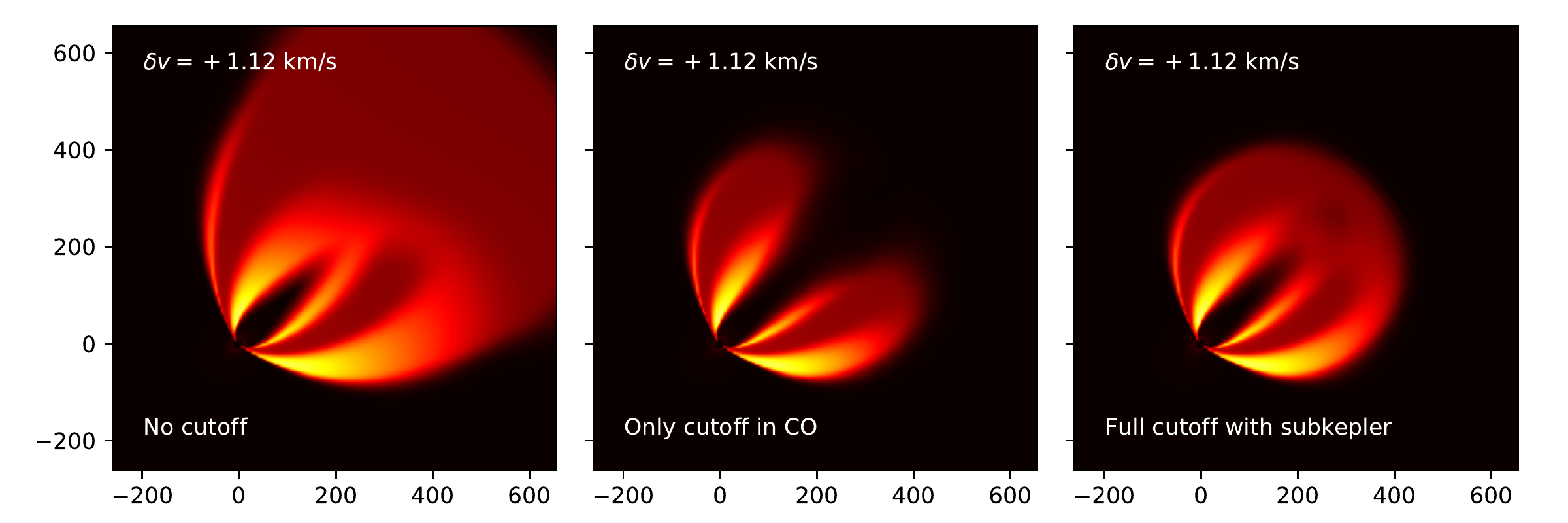}}
\caption{\label{fig-compare-subkep}Comparing the model predictions for the
  $\delta v=+1.12\,\mathrm{km/s}$ channel. Left: the original standard model
  without outer edge (or more precisely: with only the standard Lynden-Bell \&
  Pringle cutoff of $\eta=1$). Middle: the same model as left for the
  H$_2$+He gas, but a $\eta=2$ edge for the CO gas. Right: Both
  H$_2$+He and CO gas outer edge with $\eta=2$, i.e.\ with the resulting
  strong deviations from Keplerian rotation.}
\end{figure*}

To verify that the normal disk without an outer edge ($\eta=1$) is not
consistent with the data, we compare, in Fig.~\ref{fig-compare-subkep}, the
predictions of channel $\delta v=+1.12\,\mathrm{km/s}$ between three models: The
original Lynden-Bell \& Pringle model with only a very weak cut-off of $\eta=1$
(we call this ``no edge''); a model with an outer edge ($\eta=2$) for the
CO, but not for the H$_2$+He gas (thus still close to keplerian rotation); and a
model with an $\eta=2$ outer edge for both the CO and the H$_2$+He gas (the model
shown in Fig.~\ref{fig-model-channels-yes-cutoff}), and hence a strongly
subkeplerian rotation.

This comparison shows that the model without outer edge (only a weak
$\eta=1$ tapering according to Lynden-Bell \& Pringle; left panel of
Fig.~\ref{fig-compare-subkep}) does not match the data at all. There is far too
much CO emission at large radii ($r\gtrsim 500\,\mathrm{au}$). This is, in
itself, not surprising. From the zero-moment maps it was already known before
that the CO emission suddenly drops beyond $r\gtrsim 500\,\mathrm{au}$
\citep[e.g.][]{2018ApJ...869L..49I}. The middle panel of
Fig.~\ref{fig-compare-subkep} is meant to verify if an $\eta=2$ exponential
cut-off only in the CO abundance, but not in the pressure-providing H$_2$+He
gas, could explain the lack of emission at large radii. In this case the
pressure gradient remains the same as in the standard model, and thus the disk
rotates nearly keplerianly. As one can see in the figure, indeed the reduced CO
abundance at large radii prevents too much emission at large radii, consistent
with the data. But the ``teacup handle'' is not closed, which is inconsistent
with the data. Only when we apply the $\eta=2$ cut-off to the CO and the
H$_2$+He (right panel of Fig.~\ref{fig-compare-subkep}) we see that the ``teacup
handle'' is closed, as in the observed data.

The evidence for this subkeplerian rotation in the outer disk regions is,
admittedly, not conclusive. In the outer regions of the disk ($r\gtrsim
200\,\mathrm{au}$), from where this evidence is derived, the signal-to-noise
becomes sub-optimal, so we cannot exclude other effects that could explain the
shape of the low-velocity channel maps.

But if the evidence holds up, then this would indicate that the outer disk has
an outer cut-off that is sharper than predicted from viscous disk spreading,
i.e.\ the disk has an outer edge. There can be many reasons for this. One may be
that the viscous disk model is not a good description of the evolution of this
protoplanetary disk. Another one is that a process such as stellar-FUV driven
outside-in photoevaporation \revised{by the central star} \citep{2009ApJ...705.1237G} may be truncating the
disk. Or there is an unseen companion that is massive enough to truncate the
disk from the outside, but not massive enough to be detected through direct
imaging.  Finally, it could also be that the disk's turbulent viscosity is so
low, that the disk has not yet had time to viscously expand.

\section{Effect of freeze-out of CO}
\label{sec-freezeout}

\revised{
The method of measuring the midplane gas temperature described in Sections
\ref{sec-radtrans-model} and \ref{sec-comaps-hd163296}
relies on the $^{12}$CO $J=2-1$ line being optically thick even at the cold
midplane. If CO freeze-out onto dust grains does not play a role, then this
condition is easily met. However, there is evidence that CO freeze-out is taking
place in the disk around HD 163296. The most compelling evidence is the
detection of ring-shaped N$_2$H$^{+}$ $J=3-2$ emission
\citep{2015ApJ...813..128Q}. In fact, for any disk in which the midplane
temperature drops below $\sim 20\,\mathrm{K}$ in the outer regions, CO freeze-out
is likely to occur. This could drastically lower the optical depth of the
gas-phase CO in the $J=2-1$ line in the midplane regions.
}

\revised{
In fact, for HD 163296 our measurements of Section \ref{sec-comaps-hd163296}
yield midplane temperatures that are not far from the
temperature at which CO molecules tend to freeze out onto dust grains. It is
therefore not unlikely that (a) part of the gaseous CO is frozen out (and its
gas-phase abundance thus reduced), and consequently that (b) the temperature we
measure is, in fact, the temperature of the ``CO snow \revisedagain{surface}''. 
}

\revised{
  The ``CO snow \revisedagain{surface}'' is \revisedagain{an axisymmetric curved surface}.
  At the midplane,
where the temperature at a given radius is the lowest, the location of this snow
\revisedagain{surface (defined at the midplane as the snow line)} can be relatively far in,
down to perhaps even \revisedagain{75} au
\revisedagain{\citep[][when correcting for the new Gaia distance]{2015ApJ...813..128Q}}. But at the surface the gas is warmer, and the CO
\revisedagain{stays in the gas phase} much further out.  It might therefore be that, while the surface
emission remains unaffected by CO freeze-out, the midplane CO
emission/extinction may be strongly suppressed. In the most extreme case it even
absent altogether. In that case, our method of measuring the gas temperature by
employing Eddington-Barbier will, instead, measure the gas temperature at
the bottom of the non-frozen-out surface layer on the far side \citep[see
  e.g.\ the illustration in Fig.~3 of][]{2018A&A...609A..47P}. Since that is, by
definition, the freeze-out temperature, we would be measuring the freeze-out
temperature as a function of radius, instead of the midplane temperature.
This, by itself, is not uninteresting, because it allows us to calibrate
the CO freeze-out temperature under different conditions. In fact, 
it is dependent on gas pressure
\citep[e.g.][]{2005ApJ...620..994D}, so it is not surprising that we find a
shallow, but gradual decrease of the temperature with radius
(cf.\ Fig.~\ref{fig-obs-temp-mid-bins-err}).
}

\revised{
If we wish to measure the midplane temperature, and not the freeze-out
temperature, the question is: how strongly must the CO freeze-out deplete the gas-phase CO
to start affecting the reliability of our method? For our model the $^{12}$CO
$J=$2-1 line is very optically thick (of the order of $\tau\sim 10^3$ or higher,
see Fig.~\ref{fig-ear-3d-tau}). A
factor 5 or 10 reduction in the gas-phase CO abundance, such as suggested
by \citet{2015ApJ...813..128Q}, would still leave the midplane optically
thick at line-center, as required for our method. But what if the depletion
is more extreme?
}

\begin{figure*}
\centerline{\includegraphics[width=0.95\textwidth]{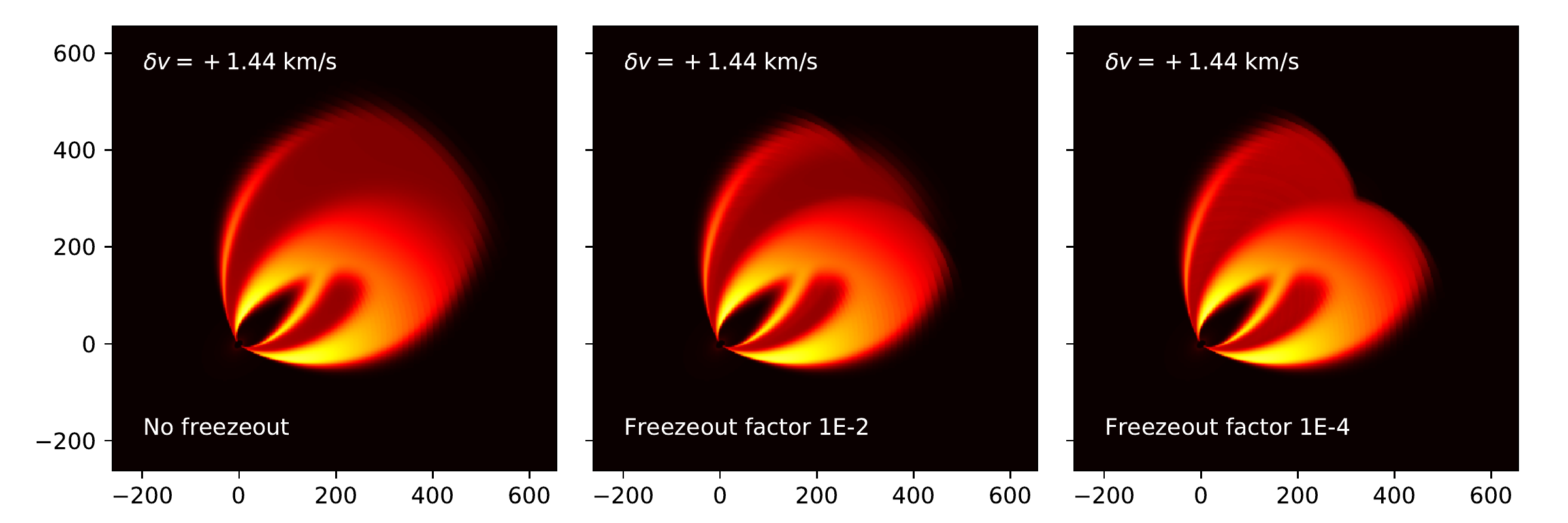}}
\caption{\label{fig-compare-freezeout}Comparing the model predictions for the
  $\delta v=+1.44\,\mathrm{km/s}$ channel for different levels of CO
  freeze-out. Left: No freeze-out, middle: freeze-out by a factor of hundred,
  right: freeze-out by a factor of ten-thousand.}
\end{figure*}

\revised{
To test this, we adjust our model in the following way. We set the CO freeze-out
temperature at $T_{\mathrm{freezeout}}=23\,\mathrm{K}$, so that the midplane
freeze-out radius lies at about 95 au.
We now mimick the freeze-out by reducing the CO abundance by a factor
$f_{\mathrm{CO}}$ everywhere $T<T_{\mathrm{freezeout}}$. We experiment with
several values of $f_{\mathrm{CO}}$. In Fig.~\ref{fig-compare-freezeout} we show
the resulting channel maps for $\delta v=1.44\,\mathrm{km/s}$, for three values
of $f_{\mathrm{CO}}$: $f_{\mathrm{CO}}=1$ (left), $f_{\mathrm{CO}}=10^{-2}$
(middle) and $f_{\mathrm{CO}}=10^{-4}$ (right).
}

\revised{
By comparing the left and the middle panel of Fig.~\ref{fig-compare-freezeout},
one can see that a factor of 100 reduction of gas-phase CO near the midplane
still leaves the midplane region optically thick. However, an optically thin
layer is formed between the midplane and the warm surface layer. One can see
through this layer, and the slightly warmer temperature of the bottom of this
layer is revealed. In the right panel, with a factor of 10000 reduction of
gas-phase CO, the entire region between the warm surface layers is optically
thin. The image looks like an empty ``oyster shell''. In this case, one is
looking at the bottom of the warm surface layer, as suggested in
\citet{2018A&A...609A..47P}.
}

\begin{figure*}
\centerline{\includegraphics[width=0.95\textwidth]{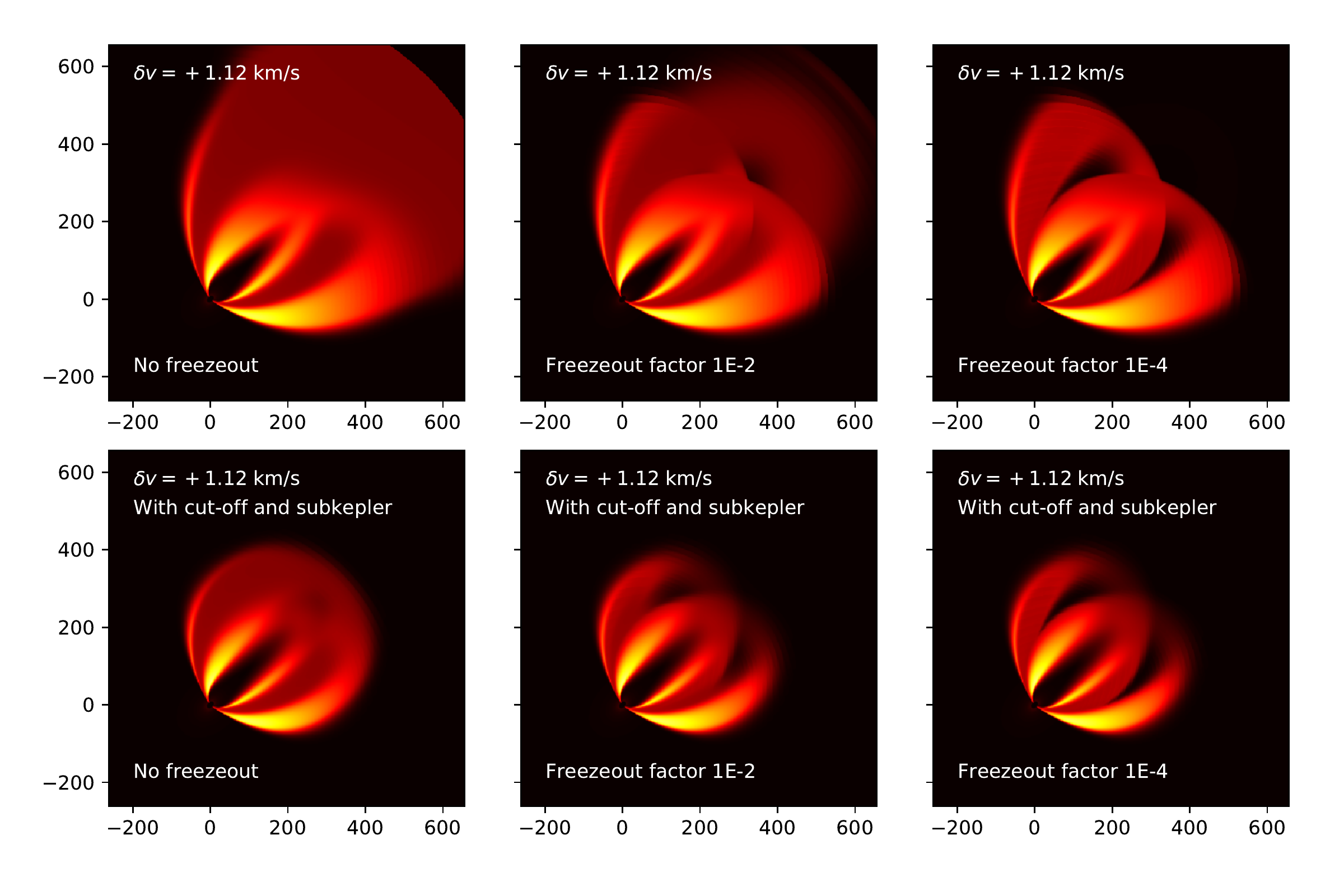}}
\caption{\label{fig-compare-freezeout-cutoff}Like Fig.~\ref{fig-compare-freezeout},
  but now for a lower velocity channel ($\delta v=+1.12\,\mathrm{km/s}$),
  thus emphasizing the outer disk regions. Compared are the standard
  model, without the outer disk cut-off ($\eta=1$) in the top row, and the
  model with a steeper disk cut-off ($\eta=2$) and resulting strong sub-Kepler
  rotation in the bottom row. Note that this is the same velocity channel as used in
  Fig.~\ref{fig-compare-subkep}, and the leftmost/rightmost panel of that figure
  are the same as the top/bottom-left panel here.}
\end{figure*}

\revised{
The effect of the freeze-out is perhaps somewhat subtle in the
$\delta v=+1.44\,\mathrm{km/s}$ channel shown in Fig.~\ref{fig-compare-freezeout}.
It becomes more pronounced in the $\delta v=1.12\,\mathrm{km/s}$ channel,
as shown in Fig.~\ref{fig-compare-freezeout-cutoff}, because lower-velocity
channels emphasize the very outer regions more. In this figure we also
compare the standard disk model ($\eta=1$) to the cut-off disk model ($\eta=2$).
The $eta=1$ models clearly produce too much emission at large radii, inconsistent
with the observations. Also the geometry for the frozen-out cases does not
seem to match the observations. The $\eta=2$ model match the lack of emission
at very large radii. The freeze-out cases (bottom-middle and bottom-right panels)
produce a geometry reminiscent of a ``tulip flower''. This is caused by the
detachment of the top and bottom surface layers, because the gas in between is
no longer optically thick. While a more detailed comparison to the observations
(using an ALMA simulator) would be necessary to rule this geometry out, a simple
by-eye comparison to the lower-left panel of Figs.~\ref{fig-obs-chanmap-mid},
\ref{fig-channels_with_kepcurves} reveals no clear evidence of this
``tulip flower'' geometry.
}

\revised{
These results depend not only on the CO depletion factor $f_{\mathrm{CO}}$, but
also on the disk mass. \citet{2019ApJ...882L..31B} recently found evidence that
the gas mass of HD 163296 is $M_{\mathrm{disk}}=0.31\,M_{\odot}$, which is about
ten times more massive as the model of this paper. Given that our model assumes
LTE, increasing the number density of CO everywhere by that same factor of ten,
would require a ten times stronger depletion to obtain the same results for the
disk's optical depth in $^{12}$CO $J=2-1$ as in
Fig.~\ref{fig-compare-freezeout}. 
}

\revised{
Astrochemical models such as those of
\citet{2006ApJ...642.1152A} and \cite{2010ApJ...722.1607W} predict gas phase CO
depletion factors as strong as $f_{\mathrm{CO}}\sim 10^{-4}$, in which case, if
we probe beyond the CO snowline radius, our method likely measures the
freeze-out temperature, not the midplane temperature (at \revisedagain{least} for the cold outer
regions).
}

\revised{
On the other hand, in many protoplanetary disks, including HD 163296, the outer
regions of the disk are likely to have a strongly reduced dust content, as a
result of radial drift of dust grains \citep{2014ApJ...780..153B}. Indeed, for
HD 163296 the dust thermal emission is confined to the inner 200 au, with most
of the emission coming from within 120 au. With less dust surface area to freeze
on to, the time scale for freeze-out increases. Turbulent mixing subsequently
could increase the abundance of gas-phase CO in the deep interiors of the disk
\citep{2017ApJ...835..162X}. Whether this effect is enough to make the midplane
optically thick in the $^{12}$CO $J=$2-1 line is unclear and requires further
modeling.
}

\revised{
In conclusion: the possibility that CO freeze-out could render the $J=2-1$ line
optically thin between the two warm surface layers of the disk, could jeopardize
the efficacy of our method of measuring the midplane gas temperature. For disks
with even higher inclination it will be possible to directly check if the gas
between the warm surface layers is optically thick or thin: In the case of
optically thin, we should see the back-side and front-side emission detach
from each other, since we will then be able to look through this otherwise
optically thick region.
}

\section{Conclusions}
In this paper we analyzed the CO channel maps of HD 163296 in more detail
than in our earlier work. We find:
\begin{enumerate}
\item By employing the Eddington-Barbier rule to line channel maps of
  sufficiently inclined protoplanetary disks, it is possible to directly measure
  the midplane gas temperature of the disk from the channel maps, without having
  to do model fitting. \revised{The only model assumption is that the line is
    very optically thick, also near the midplane. Strong freeze-out could
  invalidate this assumption.} The method requires the vertical structure of the CO
  emission to be spatially resolved well, so the method lends itself best for
  the outer regions of the disk.
\item For HD 163296 we find the midplane temperature to be almost flat
  around 20 K. There is a gentle slope, from $\sim$25 K at 100 au down to
  $\sim$18 K at 400 au.
\item If the observations do not show emission between the ``dragon fly wings'',
  our method can provide an upper limit on the midplane temperature, if the
  signal-to-noise ratio and beam smearing issues are properly accounted for. 
\item The fact that, for HD 163296, the temperature we find is close to the CO
  freeze-out temperature could mean that we, in fact, measure the temperature
  right at the vertical CO freeze-out surface, \revised{as suggested for IM Lupi in
  \citet{2018A&A...609A..47P}. If the CO freeze-out factor is only moderate
  (a factor of 5 or 10), as in \citet{2015ApJ...813..128Q}, the midplane would
  still be optically thick. But chemical models tend to predict much stronger
  freeze-out, which would yield the midplane optically thin. Our analysis does
  not give conclusive evidence for either scenario.}
\item We find tentative evidence for strong subkepler motion in the very outer
  regions of the disk around HD 163296. This is consistent with a steep negative
  pressure gradient there, suggesting that the outer edge seen in the CO
  emission is not just the removal (photodissociation or freezeout) of CO, but
  the outer edge of the gaseous disk itself.
\end{enumerate}

\begin{acknowledgements}
  C.P.D.\ thanks Rich Teague for useful discussions, and for a preview to the
  results of his inference of $v_\phi$ from the same channel map dataset.
  C.P.D.\ acknowledges work presented by Alex Fontana in his Masters thesis that
  partly inspired the work presented in this paper. C.P.D.~and N.D.~acknowledge
  funding by the DFG under grant number DU 414/20-1, which is part of SPP 1992
  ``Diversity of Extrasolar Planets''. S.A.\ acknowledges support from the
  National Aeronautics and Space Administration under Grant No.\ 17-XRP17\_2-002
  issued through the Exoplanets Research Program.
  We thank the anonymous referee for an important and useful
  suggestion that has helped us improve the paper.
\end{acknowledgements}

\begingroup
\bibliographystyle{aa}
\bibliography{ms}
\endgroup

\appendix

\section{Effect of the channel width}
\label{app-channel-width}

The high-resolution CO ALMA observations from the DSHARP campaign have a channel
width of 0.32 km/s. This is not negligible compared to the intrinsic line width.
In fact, when one overplots the channel maps of two neighboring channels, one notices
that the ``teacup handles'' are hardly overlapping, meaning that we are clearly on the border
of unresolving the line in frequency-space. Instead of the DSHARP data, we could use
the earlier data from \citet{2016PhRvL.117y1101I}, since these data have higher
velocity resolution, but they have lower spatial resolution and less sensitivity.

In this appendix we check, using the {\small\tt RADMC-3D} model, how large the
deviation from the true midplane temperature is, when applying the
Eddington-Barbier rule to a channel with width 0.32 km/s. To this end we repeat
the experiment we did in Section \ref{sec-radtrans-model}. But now we make 32
``subchannel'' images between $\delta v=1.76\pm 0.16\,\mathrm{km/s}$, i.e.\ in
total over a channel width of 0.32 km/s. We then take the mean of all 32
subchannel images and use that to extract the midplane temperature.

\begin{figure*}\centerline{
    \includegraphics[width=0.95\textwidth]{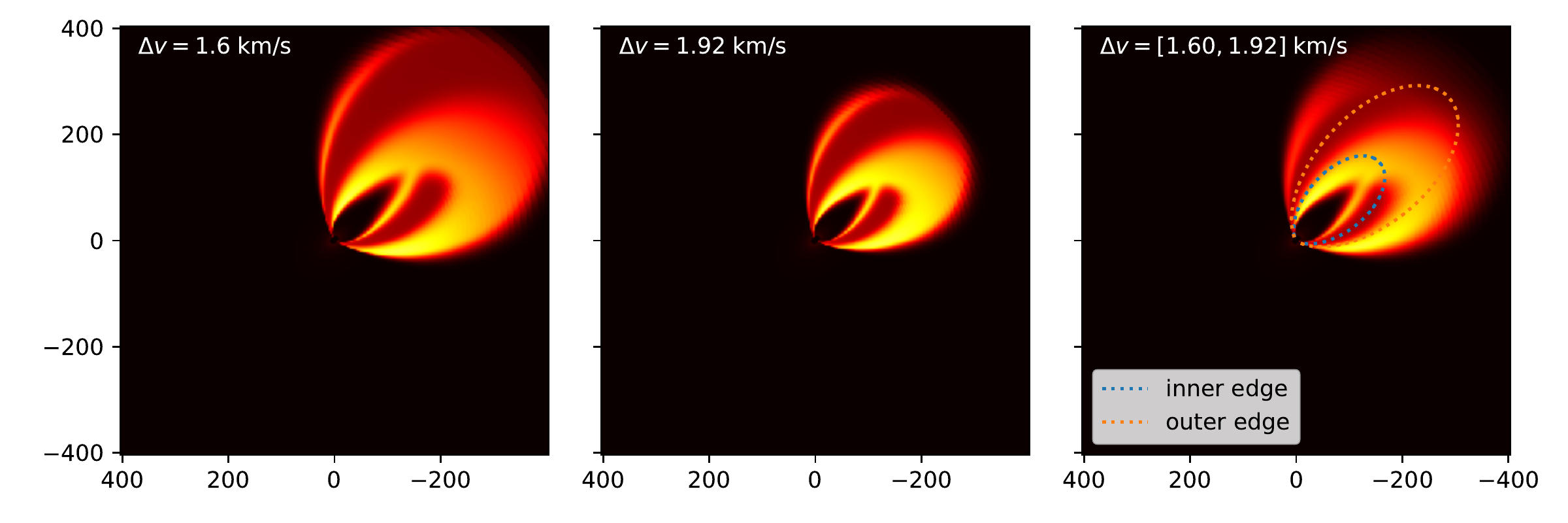}
}
\caption{\label{fig-ear-3d-wideband}The CO 2-1 channel map, like
  Fig.~\ref{fig-ear-3d}, but now for a finite-width channel with width 0.32 km/s
  (see appendix \ref{app-channel-width}). Left: the zero-width channel map at
  $\delta v=1.76-0.16\;\mathrm{km/s}$. Middle: the zero-width channel map at
  $\delta v=1.76+0.16\;\mathrm{km/s}$. These are the two extreme sides of the
  0.32 km/s band: Left the low velocity extreme, middle the high velocity
  extreme. In between another 30 intermediate ``subchannel'' maps are
  computed. The mean of all these subchannel maps is shown to the right.}
\end{figure*}

\begin{figure*}
\centerline{\includegraphics[width=0.5\textwidth]{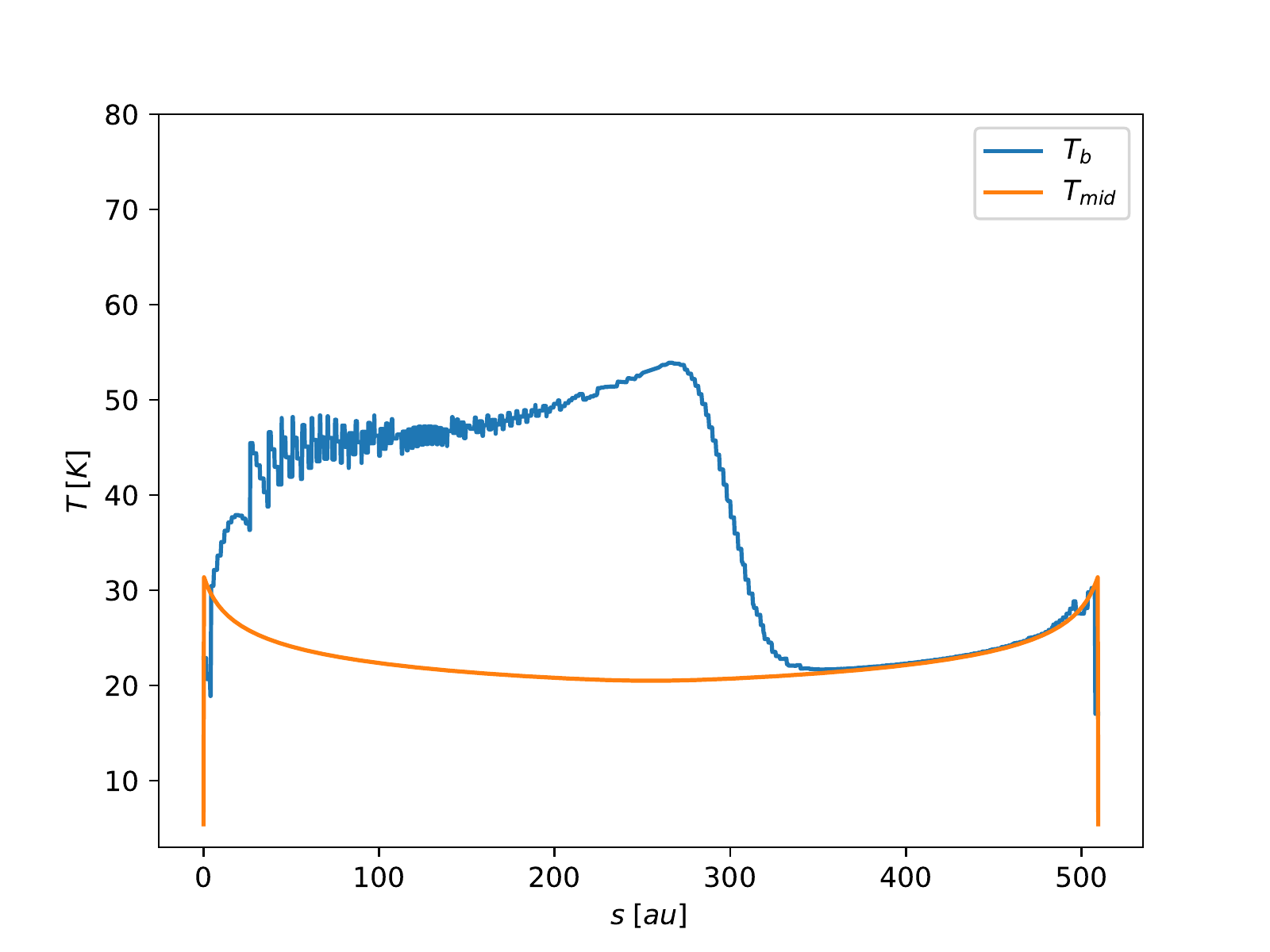}
\includegraphics[width=0.5\textwidth]{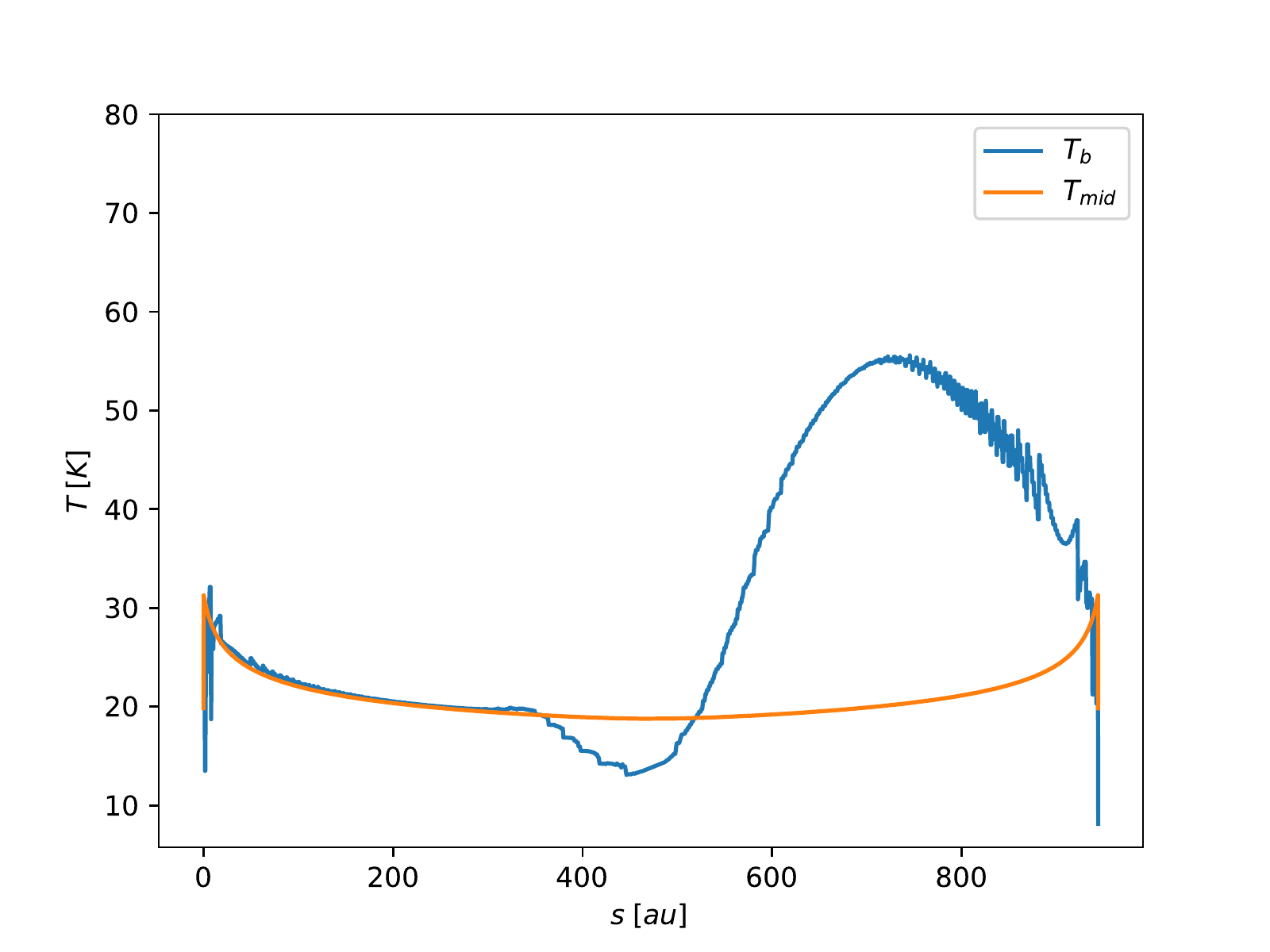}}
\caption{\label{fig-wideband-paths}As Fig.~\ref{fig-paths}, but now for the band-averaged
  channel map model of Fig.~\ref{fig-ear-3d-wideband}-right. Left:
  the inner path. Right: the outer path. Overplotted is the actual midplane
  temperature of the model.}
\end{figure*}

In Fig.~\ref{fig-ear-3d-wideband} the effect of the band-averaging is shown.
The shape of the averaged ``teacup handle'' (panel on the right) is a bit
fuzzier than the original single-frequency maps. But overall the shape remains
the same.  The resulting Eddington-Barbier midplane temperature extractions are
shown in Fig.~\ref{fig-wideband-paths}. While the match between the extraction
(blue curve) and the real midplane temperature (orange curve) is less good than
for the single-frequency case (Fig.~\ref{fig-paths}), the error is only about 1
to 2 Kelvin. This is well within the noise of the data of HD 163296.

\section{Effect of unresolved turbulence}
\label{app-turbulence}

\begin{figure}
\centerline{\includegraphics[width=0.5\textwidth]{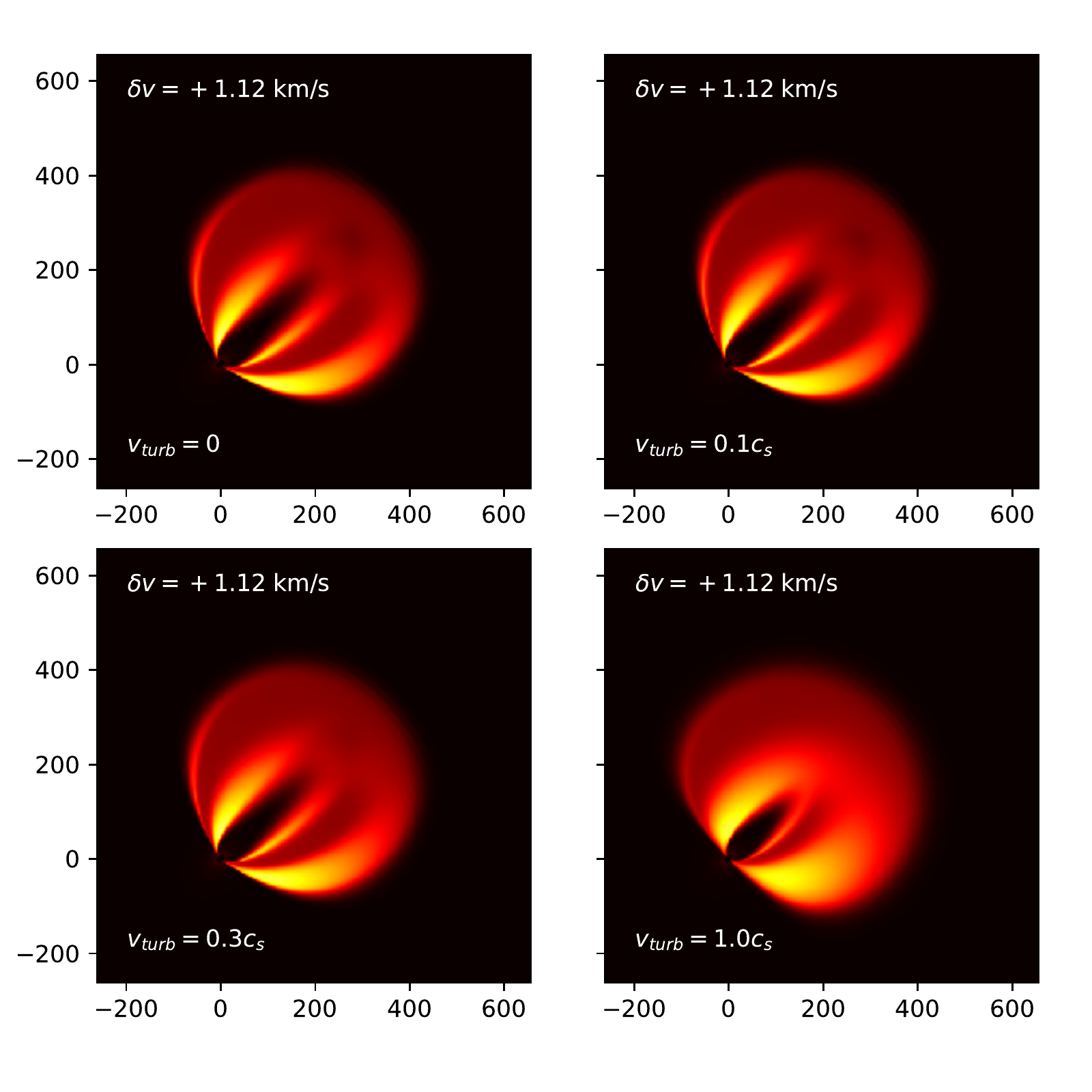}}
\caption{\label{fig-compare-turb}Demonstration of the effect of unresolved
  turbulence in the disk on the $\delta v=1.12\,\mathrm{km/s}$ channel map.
  Only for very strong turbulence ($v_{\mathrm{turb}}=c_s$ do the effects
  become so large that they are easily seen in the maps. This is why it
  is so challenging to measure the strength of turbulence from the
  channel maps, as studied by \citet{2015ApJ...813...99F}.}
\end{figure}

In the standard models of this paper we set the unresolved turbulence to zero.
The only line broadening is due to the thermal broadening of the CO molecules
themselves. Here we demonstrate the effect of possible turbulent broadening
on the shape of the ``teacup handle''. We start from the subkepler ($\eta=2$)
model of Section \ref{sec-subkep}. We then add turbulence at three different
magnitudes compared to the local sound speed: $v_{\mathrm{turb}}=0.1\,c_s$,
$v_{\mathrm{turb}}=0.3\,c_s$, and $v_{\mathrm{turb}}=1\,c_s$. The results are
shown in Fig.~\ref{fig-compare-turb}.

\section{2-D disk structure and orbital velocity}
\label{sec-twodee-structure}

Due to the radial gas pressure gradient, the protoplanetary disk does not orbit
with exact Kepler velocity. The vertical gas pressure gradient also affects the
orbital velocity of the gas, albeit in an indirect way. In cylindrical
coordinates it is possible to set up an axially symmetric gas disk structure in
which the gravity from the star, the gas pressure gradients in the disk and the
centrifugal force are in exact balance \citep{2013ApJ...774...16R}. Let us
define $r_{\mathrm{cyl}}$ to be the cylindrical radial coordinate, while $r_{\mathrm{spher}}$ is the
spherical radial coordinate. $z$ is the vertical coordinate measured from the
midplane of the disk. They are related via $r_{\mathrm{spher}}^2=r_{\mathrm{cyl}}^2+z^2$. The gravitational
force per unit mass of gas is
\begin{equation}
{\bf f}_g = -\frac{GM_{*}}{r_{\mathrm{spher}}^3}\left(r_{\mathrm{cyl}}{\bf e}_{r_{\mathrm{cyl}}}+z{\bf e}_{z}\right)
\end{equation}
The centrifugal force is
\begin{equation}
{\bf f}_c = \frac{v_\phi^2}{r_{\mathrm{cyl}}}{\bf e}_{r_{\mathrm{cyl}}}
\end{equation}
and the pressure gradient force is
\begin{equation}
  {\bf f}_p = - \frac{1}{\rho}\frac{\partial p}{\partial r_{\mathrm{cyl}}}{\bf e}_{r_{\mathrm{cyl}}}
  - \frac{1}{\rho}\frac{\partial p}{\partial z}{\bf e}_{z}
\end{equation}
where $p=\rho c_s^2$ is the gas pressure, with $c_s$ the isothermal
sound speed. Force balance is reached when
${\bf f}_g+{\bf f}_c+{\bf f}_p=0$.  For the ${\bf e}_{z}$ component this leads to
the equation
\begin{equation}\label{eq-vert-equil}
\frac{\partial \ln p(r_{\mathrm{cyl}},z)}{\partial z} = -\frac{GM_{*}}{(r_{\mathrm{cyl}}^2+z^2)^{3/2}}\frac{z}{c_s(r_{\mathrm{cyl}},z)^2}
\end{equation}
The right-hand-side does not depend on $p$ or $\rho$, but does depend on $z$.
If the temperature structure, and thus $c_s(r_{\mathrm{cyl}},z)$, is a {\em given} function of $r_{\mathrm{cyl}}$ and $z$, then
Eq.~(\ref{eq-vert-equil}) can be directly numerically integrated from the
midplane upward, assuming we know the midplane pressure $p(r_{\mathrm{cyl}},0)$. This yields
the vertical pressure profile $p(r_{\mathrm{cyl}},z)$. By dividing this by the known
$c_s(r_{\mathrm{cyl}},z)^2$, we obtain the density profile $\rho(r_{\mathrm{cyl}},z)$. Typically we do not
know $p(r_{\mathrm{cyl}},0)$ a-priori, but instead we prescribe the radial surface density
profile
\begin{equation}\label{eq-surfdens-def}
\Sigma(r_{\mathrm{cyl}}) = \int_{-\infty}^{+\infty}\rho(r_{\mathrm{cyl}},z)dz
\end{equation}
Numerically we start the vertical integration of Eq.~(\ref{eq-vert-equil}) with
a trial value of $p(r_{\mathrm{cyl}},0)$, integrate Eq.~(\ref{eq-vert-equil}), and rescale
$\rho(r_{\mathrm{cyl}},z)$ to match Eq.~(\ref{eq-surfdens-def}). This immediately gives us
the 2-D density structure.

Next we demand that also the radial forces are in balance. This yields
the equation
\begin{equation}\label{eq-radial-force-balance}
\frac{v_\phi^2}{r_{\mathrm{cyl}}}-\frac{GM_{*}}{r_{\mathrm{spher}}^3}r_{\mathrm{cyl}}-\frac{1}{\rho}\frac{\partial p}{\partial r_{\mathrm{cyl}}} = 0
\end{equation}
Although this looks like a differential equation, it is not, because we already
know the 2-D structure of $p(r_{\mathrm{cyl}},z)$. So we can numerically evaluate the third
term of the equation, as well as the second one. This immediately leads to the
value of $v_\phi(r_{\mathrm{cyl}},z)$. The resulting solution is in perfect 2-D force
balance.  It can happen that Eq.~(\ref{eq-radial-force-balance}) is not
solvable, i.e.  yields imaginary $v_\phi$ velocity. In that case the radial
pressure gradient force is apparently larger than the gravitational force, in
which case no azimuthal velocity can be found that leads to equilibrium. In
practice, this part of the disk will likely escape through a thermal wind.

Note that even without substantial pressure support, Eq.~(\ref{eq-radial-force-balance})
shows that the balance between centrifugal and gravitational force leads to a
reduction of $|v_{\phi}|$ with increasing distance from the midplane. The surface
layers of a disk therefore tend to rotate slower than the midplane
\citep[see e.g.][]{2013ApJ...774...16R}.

\section{Simple temperature prescription}
\label{sec-temp-prescription}

To keep the radiative transfer model of Section \ref{sec-radtrans-model} as
simple as possible, we decided to prescribe the 2-D/3-D temperature structure of
the disk, rather than compute it. We use the measured midplane and surface
temperature radial profiles (Eq.~\ref{eq-midplane-temp} and
\ref{eq-surface-temp}) as the basis of this prescription. We use cylindrical
coordinates, and interpret the $r$ in Eqs.~\ref{eq-midplane-temp},
\ref{eq-surface-temp} as the cylindrical radius $r_{\mathrm{cyl}}$. We parameterize the
$z$-dependence of the temperature $T(r_{\mathrm{cyl}},z)$ as
\begin{equation}
T(r_{\mathrm{cyl}},z) = \left\{T^4_{\mathrm{mid}}(r_{\mathrm{cyl}}) + f(z) T^4_{\mathrm{surf}}(r_{\mathrm{cyl}})\right\}^{1/4}
\end{equation}
where 
\begin{equation}\label{eq-transition-mid-surf}
f(z) = \frac{1}{2} \tanh\left(\frac{z/r_{\mathrm{cyl}}-h_s}{w_s}\right)+\frac{1}{2}
\end{equation}
This introduces two parameters: the dimensionless height of the surface layer
$h_s$ and the dimensionless width of the transition from midplane to surface
temperature $w_s$. Eq.~(\ref{eq-transition-mid-surf}) creates a smooth two-layer
disk model, with $T=T_{\mathrm{mid}}$ below the surface layer and
$T=T_{\mathrm{mid}}$ above. 

\end{document}